\documentclass[conference]{IEEEtran}

\usepackage{amsmath,amsfonts}
\usepackage{stmaryrd}
\usepackage{algorithm}
\usepackage{algpseudocodex}
\usepackage{tcolorbox}
\usepackage{threeparttable}
\usepackage{graphicx}
\usepackage{bbding} 
\usepackage{tikz}
\newcommand*\emptycirc[1][1ex]{\tikz\draw (0,0) circle (#1);} 
\newcommand*\halfcirc[1][1ex]{%
	\begin{tikzpicture}
	\draw[fill] (0,0)-- (90:#1) arc (90:270:#1) -- cycle ;
	\draw (0,0) circle (#1);
	\end{tikzpicture}}
\newcommand*\fullcirc[1][1ex]{\tikz\fill (0,0) circle (#1);} 
\newcommand*\redcirc[1][1ex]{\tikz\fill[red] (0,0) circle (#1);} 
\newcommand*\greencirc[1][1ex]{\tikz\fill[green] (0,0) circle (#1);} 
\newcommand*\bluecirc[1][1ex]{\tikz\fill[blue] (0,0) circle (#1);} 
\newcommand{\bluecheck}{{\Checkmark}}
\newcommand{\redXsolid}{{\XSolidBrush}}

\usepackage[font=small, justification=centerlast]{subfig}

\usepackage{cite}
\usepackage{pifont}
\usepackage{xurl}
\usepackage{tabularx}
\usepackage{tabularray}
\usepackage{hyperref}  
\hypersetup{
    colorlinks=true,
    linkcolor=black,
    filecolor=magenta,      
    urlcolor=cyan,
    citecolor=red,
    pdftitle={Overleaf Example},
    pdfpagemode=FullScreen,
    }

\usepackage[nameinlink]{cleveref}
\crefname{figure}{\figurename}{\figurename}
\crefname{table}{\tablename}{\tablename}
\makeatletter
\crefname{algorithm}{\ALG@name}{\ALG@name}
\makeatother
\crefname{equation}{Eq.}{Eq.}
\crefname{section}{Section}{Section}
\crefname{subsection}{Section}{Section}
\crefname{subsubsection}{Section}{Section}
\crefname{definition}{Definition}{Definition}
\crefname{theorem}{Theorem}{Theorem}
\crefname{remark}{Remark}{Remark}
\DeclareMathAlphabet{\mathtt}{OT1}{cmtt}{m}{n}

\newtheorem{theorem}{Theorem}
\newtheorem{Define}{Definition}

\usepackage{hyperref}
\hypersetup{
    colorlinks=true,
    linkcolor=blue,
    filecolor=magenta,      
    urlcolor=cyan,
    pdftitle={Overleaf Example},
    pdfpagemode=FullScreen,
    }

\graphicspath{{figure/}{bio-figure/}}

\begin{document}

\title{\huge Kangaroo: A Private and Amortized Inference Framework over WAN for Large-Scale Decision Tree Evaluation}

\author{%
\IEEEauthorblockN{%
Wei Xu\IEEEauthorrefmark{1},
Hui Zhu\IEEEauthorrefmark{1}\textsuperscript{\ding{41}},
Yandong Zheng\IEEEauthorrefmark{1},
Song Bian\IEEEauthorrefmark{2},
Ning Sun\IEEEauthorrefmark{1},
Hao Yuan\IEEEauthorrefmark{1},
Dengguo Feng\IEEEauthorrefmark{3}, and 
Hui Li\IEEEauthorrefmark{1}}
\IEEEauthorblockA{\IEEEauthorrefmark{1}%
Xidian University, \{xuwei\_1,sunning,yuan\_hao\}@stu.xidian.edu.cn,\\\{zhuhui,zhengyandong\}@xidian.edu.cn,lihui@mail.xidian.edu.cn}
\IEEEauthorblockA{\IEEEauthorrefmark{2}%
Beihang University, sbian@buaa.edu.cn}
\IEEEauthorblockA{\IEEEauthorrefmark{3}%
School of Cyber Science and Technology, fengdg@263.net}
}


\IEEEoverridecommandlockouts
\makeatletter\def\@IEEEpubidpullup{6.5\baselineskip}\makeatother
\IEEEpubid{\parbox{\columnwidth}{
    Network and Distributed System Security (NDSS) Symposium 2026\\
    23 - 27 February 2026 , San Diego, CA, USA\\
    ISBN 979-8-9919276-8-0\\
    https://dx.doi.org/10.14722/ndss.2026.230892\\
    www.ndss-symposium.org
}
\hspace{\columnsep}\makebox[\columnwidth]{}}

\maketitle

\begin{abstract}
	With the rapid adoption of Models-as-a-Service, concerns about data and model privacy have become increasingly critical. To solve these problems, various privacy-preserving inference schemes have been proposed. In particular, due to the efficiency and interpretability of decision trees, private decision tree evaluation (PDTE) has garnered significant attention. However, existing PDTE schemes suffer from significant limitations: their communication and computation costs scale with the number of trees, the number of nodes, or the tree depth, which makes them inefficient for large-scale models, especially over WAN networks. To address these issues, we propose Kangaroo, a private and amortized decision tree inference framework build upon packed homomorphic encryption. Specifically, we design a novel model hiding and encoding scheme, together with secure feature selection, oblivious comparison, and secure path evaluation protocols, enabling full amortization of the overhead as the number of nodes or trees scales. Furthermore, we enhance the performance and functionality of the framework through optimizations, including same-sharing-for-same-model, latency-aware, and adaptive encoding adjustment strategies. Kangaroo achieves a $14\times$ to $59\times$ performance improvement over state-of-the-art (SOTA) one-round interactive schemes in WAN environments. For large-scale decision tree inference tasks, it delivers a $3\times$ to $44\times$ speedup compared to existing schemes. Notably, Kangaroo enables the evaluation of a random forest with $969$ trees and $411825$ nodes in approximately $60$ ms per tree (amortized) under WAN environments.
\end{abstract}

\IEEEpeerreviewmaketitle

\section{Introduction}
With the rapid development of machine learning technologies, ``Model as a Service" (MaaS) has been widely applied across various domains~\cite{GanWY23,BarcenaDMR25,MLapplication}. By deploying models on the server, clients can benefit from convenient and efficient services. Although the approach improves service efficiency, clients are required to upload sensitive personal data, such as medical diagnostic records, financial transaction details, and personal communication information, to the server, which increases the risk of privacy leakage. In contrast, running the model on the client side can effectively protect client's privacy. However, since the models hold significant intellectual property, service providers are often reluctant to share it. Therefore, achieving Model as a Service while ensuring both the client's data privacy and the server's model security has become a critical issue~\cite{HouLLWWL22,JiaGJF18}.

{Decision tree inference, as an intuitive and powerful machine learning tool, has been widely applied in various fields such as medical diagnosis, financial risk assessment, and customer behavior prediction~\cite{XGBoostCha,XGBoost,CostaP23}. Compared to deep neural networks (DNNs), decision trees and ensemble models (e.g., XGBoost, Random Forest) are often preferred in practice for their interpretability, better performance on tabular data, and ease of debugging, which are especially important in applications requiring transparency and regulatory compliance~\cite{XGBoostCha,XGBoost,CostaP23}.} To ensure both the client's data privacy and the server's model security, numerous private decision tree evaluation schemes (PDTE) have been proposed~\cite{LuZS18, Cong00P22, MahdaviNLK23, AkaviaLRRSV22, WuFNL16, TaiMZC17, KissNLAS19, BostPTG15, JoyeS18, TuenoKK19, MaT0C21, BaiSCCR22, TuenoBK20, BaiSZWCCR23, YuanLQHZX24, FuCXSLS24, ZhengDWWN22, ZhengWWDN23, LiuCLSQ20, zhang2024secure, JiZLLR23, LiuSCCSL19, ChengGMMT24}. These schemes usually consist of three steps: secure feature selection, oblivious comparison, and secure path evaluation. They leverage privacy-preserving techniques such as fully homomorphic encryption (FHE)~\cite{LuZS18, Cong00P22, MahdaviNLK23, AkaviaLRRSV22} or secure multi-party computation (MPC)~\cite{WuFNL16, TaiMZC17, KissNLAS19, BostPTG15, JoyeS18, TuenoKK19, MaT0C21, BaiSCCR22, TuenoBK20, BaiSZWCCR23, ZhengDWWN22, ZhengWWDN23, LiuCLSQ20, zhang2024secure, JiZLLR23, LiuSCCSL19, FuCXSLS24, ChengGMMT24, YuanLQHZX24} to realize the three core steps, and have been widely adopted in client-server~\cite{WuFNL16, TaiMZC17, KissNLAS19, AkaviaLRRSV22, MahdaviNLK23,BostPTG15, JoyeS18, TuenoKK19, MaT0C21, LuZS18, Cong00P22, BaiSCCR22, TuenoBK20} and outsourced scenarios~\cite{ZhengDWWN22, ZhengWWDN23, LiuCLSQ20, zhang2024secure, JiZLLR23, LiuSCCSL19, FuCXSLS24, BaiSZWCCR23, ChengGMMT24, YuanLQHZX24}.

Existing PDTE schemes offer distinct advantages and have proven effective in safeguarding both model confidentiality and client data privacy. However, these approaches have not been thoroughly evaluated for large-scale decision tree evaluation, particularly in wide-area network (WAN) environments. As task demands grow and accuracy requirements increase, decision tree models are being deployed at larger scales to enhance prediction performance and generalization~\cite{klusowski2024large,forestsprune0,breiman2001random}. Although some model pruning techniques~\cite{forestsprune1,forestsprune3,forestsprune2} have been proposed to reduce the size of decision trees, the tree depth and the total number of nodes often remain substantial. For instance, some pruned models still reach a maximum depth of $30$ and consist of up to $10^6$ nodes~\cite{forestsprune3}. These observations highlight the continued importance and relevance of large-scale decision tree evaluation~\cite{breiman2001random, forestsprune1,forestsprune3, izenman2008modern, james2013introduction}. However, in the current research on PDTE, most experimental evaluations are still limited to small-scale decision trees. This raises important questions about whether existing schemes can scale effectively to support large, deep, and complex models under real-world deployment scenarios, particularly in WAN environments with high latency and limited bandwidth, as commonly encountered in practical applications~\cite{DiaaFHDEK0LMOAK24,ZhangRPBZ22,ChilamkurthyPGC22}.

\begin{table*}
    \caption{A representative survey of private decision tree evaluation schemes for single tree.}
    \label{Summary}
    \begin{tblr}
    {
    colspec={Q[c,wd=1.3cm,font=\bfseries] c c c c c c c c c c c c c},
    columns      = {valign=m,co=-1,colsep+=-1pt},
    rows         = {halign=c},
    hline{1,Z}   = {wd=.08em},
    hline{2}     = {wd=.05em},
    row{1}       = {font=\bfseries\boldmath},
    }
                      & {Squirrel\\\cite{LuHZWH23}} & {SGBoost\\\cite{ZhaoZXWLL23}} & {FSSTree\\\cite{FuCXSLS24}} & {Cheng\\\cite{ChengGMMT24}} & {Zheng\\\cite{ZhengWWDN23}} & {Zhao\\\cite{ZhaoZWLL23}} & {Yuan\\\cite{YuanLQHZX24}} & {Ma\\\cite{MaT0C21}} & {Bai (HE)\\\cite{BaiSCCR22}} &{Tai (HHH)\\\cite{TaiMZC17}} & {Kiss (HGH)\\\cite{KissNLAS19}}  & {Sorting-Hat\\\cite{Cong00P22}} & {Levelup\\\cite{MahdaviNLK23}} & Ours \\
    Primitives        &  COT     &LHE       &RSS,FSS   &{RSS,FSS}&{SS,OT,\\TTP} &AHE   &{AHE,\\PRF}&{SS,GC,\\OT} &{AHE,SS,\\OT}&AHE       &AHE,GC    &FHE       & FHE      &PHE,SS \\   
    Round             &\greencirc&\greencirc&\bluecirc &\greencirc &\redcirc  &\greencirc&\bluecirc  &\bluecirc    &\bluecirc    &\greencirc&\greencirc&\greencirc&\greencirc&\greencirc\\
    Feature           &\emptycirc&\emptycirc&\fullcirc &\fullcirc  &\fullcirc &\halfcirc &\fullcirc  &\fullcirc    &\fullcirc    &\fullcirc &\fullcirc &\fullcirc &\fullcirc &\fullcirc\\
    Comparison        &\redXsolid&$O(\tau)$ &$O(D)$    &$O(2^D)$   &$O(2^D)$  &$O(2^D)$  &$O(D)$     &$O(D)$       &$O(D)$       &$O(\tau)$ &$O(\tau)$ &$O(\tau)$ &$O(\tau)$ &$O(1)$ \\
    Path              &COTPath   &{Poly}&OnePath&OnePath$^*$&{Poly}    &OnePath$^*$&OnePath  &OnePath      &OnePath      &Path      &Path      &{Poly}      &Path      &MixPath \\
    Sparse            &\bluecheck&\bluecheck&\bluecheck&\redXsolid &\redXsolid&\redXsolid&\bluecheck &\bluecheck   &\bluecheck   &\bluecheck&\bluecheck&\bluecheck&\bluecheck&\bluecheck\\
    OneTime           &\bluecheck&\bluecheck&\redXsolid&\redXsolid &\bluecheck&\redXsolid&\redXsolid &\redXsolid   &\bluecheck   &\bluecheck&\bluecheck&\bluecheck&\bluecheck&\bluecheck \\
    Amortized         &\bluecheck&\redXsolid&\bluecheck&\bluecheck &\bluecheck&\redXsolid&\redXsolid &\bluecheck   &\bluecheck   &\redXsolid&\bluecheck&\redXsolid&\redXsolid&\bluecheck  \\
    Scenario          &Fed       &Fed       &O-3PC     &O-3PC      &O-2PC     &O-2PC     &O-2PC      &{2PC,\\O-2PC}&2PC          &2PC       &2PC       &2PC       &2PC       &{2PC,\\O-2PC} \\
    \end{tblr}
    \par
    \setlength{\prevdepth}{0pt}
    \textsuperscript{\dag} Round: communication round, \greencirc: $O(1)$; \bluecirc: $O(D)$; \redcirc: $O(D+\log t)$; $D$: the maximum tree depth; $t$: the bit size of feature. Feature: secure feature selection, \emptycirc: no support; \halfcirc: no oblivious; \fullcirc: oblivious. Comparison: the complexity of comparison, $\tau$: the number of decision node. Path: the path evaluation, COTPath: get weight by COT, {Poly}: polynomial-based evaluation, OnePath: get weight by one path; OnePath$^*$: get weight by one path after tree permutation; Path: Path cost-based evaluation; MixPath: Combine Path with {Poly}. Sparse: sparse tree model. OneTime: one-time setup phase. Amortized: calculation amortization. \bluecheck: support; \redXsolid: no support. Fed: inference in federated learning; O: outsourcing. In the schemes~\cite{Cong00P22,MahdaviNLK23}, we only analyze the operation costs of XCMP. 
    \end{table*}

$\textbf{Motivation:}$ Based on the above questions, we analyze the current state-of-the-art techniques. According to the number of communication rounds, existing schemes can be divided into depth-round evaluation schemes and constant-round evaluation schemes. Depth-round evaluation schemes~\cite{JoyeS18, TuenoKK19, MaT0C21,BaiSCCR22, JiZLLR23, BaiSZWCCR23, YuanLQHZX24, FuCXSLS24} achieve sublinear decision tree evaluations, significantly reducing computation cost. However, as the depth of the trees increases, these schemes incur a growing number of communication rounds, leading to significant latency in WAN setting. Constant-round schemes~\cite{WuFNL16, TaiMZC17, KissNLAS19, AkaviaLRRSV22, ZhengDWWN22, BostPTG15, LiuCLSQ20, ZhengWWDN23, LuZS18, Cong00P22, MahdaviNLK23, zhang2024secure, JiZLLR23, LiuSCCSL19, ChengGMMT24,TuenoBK20} are immune to such communication cost, but their computation and communication costs grow linearly as the number of trees and decision nodes increases. As a result, the main motivation of this work is to design an efficient constant-round scheme that overcomes communication round limitations while effectively amortizing the costs introduced by large-scale tree models in WAN setting. Moreover, considering that the client-server model can be naturally extended to outsourced scenarios, this work primarily focuses on the former.

\subsection{Our Contributions}
To solve the above problems, we propose Kangaroo, a privacy-preserving and efficient constant-round inference framework for large-scale decision tree evaluation. We leverage packed homomorphic encryption (PHE) to amortize both computation and communication costs. Although some schemes have explored the use of PHE for amortization in decision tree evaluation, they fail to fully exploit its amortization potential and suffer from significant performance degradation~\cite{Cong00P22, MahdaviNLK23, ShinCLKL24}. In contrast, Kangaroo introduces a novel set of single-instruction-multiple-data (SIMD) PHE protocols over encoded models, which fully leverage the amortization capability of PHE while significantly improving inference efficiency. The key idea is to treat each coefficient in the PHE ciphertext as a model node and combine it with secret sharing (SS) to achieve full amortization and scalability. The main contributions of this work are summarized as follows.

\begin{itemize}
    \item [$\bullet$] \textbf{\textit{A new two-party inference framework is proposed for large-scale decision tree evaluation over WAN.}} To the best of our knowledge, Kangaroo is the first scheme that achieves full amortization over PHE and is specifically tailored for efficient large-scale decision tree evaluation in WAN setting. In this framework, a novel model hiding and encoding technique is introduced, where decision tree nodes are mapped to polynomial coefficients to enable efficient amortized inference.
    \item [$\bullet$] \textbf{\textit{A novel set of secure components is designed to support efficient decision tree evaluation.}} These components include packed feature selection ($\mathtt{PackFeatureSel}$), packed oblivious comparison ($\mathtt{PackObliviousCom}$), and packed path evaluation ($\mathtt{PackPathEva}$), which securely realize the three basic steps of decision tree inference. Each component is optimized to enable effective amortization, substantially minimizing both computation and communication costs. 
    \item [$\bullet$] \textbf{\textit{Several optimization strategies are introduced to enhance the practicality of Kangaroo.}} Leveraging our model hiding and encoding technique, a same-sharing-for-same-model strategy is employed to enable real-time inference responses. Additionally, a latency-aware strategy is adopted to further optimize inference efficiency. Building upon our secure components, an adaptive encoding adjustment strategy is developed to achieve full amortization for large-scale decision tree evaluation.
\end{itemize}

$\textbf{Benchmarks:}$ We have implemented Kangaroo, and the core code is available on GitHub\footnote{https://github.com/pigeon-xw/Kangaroo}. Compared to state-of-the-art (SOTA) schemes, our secure components demonstrate superior amortization capabilities and eliminate the need for any offline preprocessing. We also conducted extensive experimental evaluations over WAN. Specifically, Kangaroo achieves a $14\times$ to $59\times$ improvement over SOTA one-round interaction schemes on small-scale datasets. For large-scale decision tree evaluations, Kangaroo outperforms existing SOTA schemes by $3\times$ to $44\times$. In evaluating a random forest consisting of $969$ trees and $411825$ nodes, Kangaroo achieves an amortized inference time of approximately $60$ ms per tree.


\subsection{Related Works}
\label{relatedwork}
We summarize several representative schemes in~\cref{Summary}. Moreover, we provide a comprehensive review of private decision tree evaluation (PDTE) and their core components: oblivious comparison and secure path evaluation. Finally, we summarize the amortization techniques commonly adopted in privacy-preserving applications.

\textbf{1) Private Decision Tree Evaluation}

$\bullet$ \textbf{Client Server Model (2PC):} Existing 2PC schemes~\cite{WuFNL16, TaiMZC17, KissNLAS19, AkaviaLRRSV22, MahdaviNLK23,BostPTG15, JoyeS18, TuenoKK19, MaT0C21, LuZS18, Cong00P22, BaiSCCR22, TuenoBK20} can be divided into depth-round evaluation schemes~\cite{JoyeS18, TuenoKK19, MaT0C21,BaiSCCR22} and constant-round schemes~\cite{LuZS18, Cong00P22, MahdaviNLK23, AkaviaLRRSV22, WuFNL16, TaiMZC17, KissNLAS19, BostPTG15,TuenoBK20}. Among depth-round evaluation schemes, the most representative work is proposed by Ma et al.~\cite{MaT0C21}, which enables evaluation by traversing only one path. At each layer, it requires only one $\binom{M}{1}$-oblivious transfer (OT), one conditional OT (COT), and one garbled circuit (GC) operation, which significantly improve the efficiency of the protocol. A limitation of this protocol is that it requires re-initialization of the decision tree in each evaluation to ensure security. In constant-round schemes, Kiss et al.~\cite{KissNLAS19} summarize a set of modular components that support the three key steos of decision tree evaluation, opening up a new perspective for the design of constant-round protocols. Subsequently, many researchers shifted their focus to one-round interaction schemes~\cite{LuZS18, Cong00P22, MahdaviNLK23, AkaviaLRRSV22}, among which Sortinghat~\cite{Cong00P22} and Levelup~\cite{MahdaviNLK23} stand out as the most representative works. Both schemes optimize the XCMP protocol~\cite{LuZS18} and propose other amortizable techniques to accelerate the comparison. 

$\bullet$ \textbf{Two-party Outsourcing Computation (O-2PC):} Inspired by the design of scheme~\cite{KissNLAS19}, many O-2PC protocols~\cite{ZhengDWWN22, ZhengWWDN23, LiuCLSQ20, LiuSCCSL19, MaT0C21, ZhaoZWLL23} achieve outsourced computation by replacing 2PC components~\cite{TaiMZC17, KissNLAS19, BostPTG15, JoyeS18, MaT0C21}. For example, Zheng et al.~\cite{ZhengWWDN23} adopt the feature selection method from~\cite{MaT0C21} and the MSB bit extraction protocol~\cite{Patra0SY21} to optimize the feature selection and comparison. Based on the above optimizations, the communication rounds for comparison in~\cite{ZhengDWWN22} are reduced from $O(t)$ to $O(\log t)$. By introducing a trusted third party (TTP), their scheme supports secure multiplication to enable polynomial-based evaluation as in~\cite{BostPTG15}, with a communication round of $O(D)$. However, to conceal the structure of the decision tree, the evaluation must be performed over a complete binary tree. Zhao et al.~\cite{ZhaoZWLL23} adopt the private data comparison protocol~\cite{ZhengZLGZWSL23} to perform decision tree evaluation. In this approach, the cloud only decrypts along a single path to obtain a randomized traversal result, with the path protected by a tree permutation mechanism~\cite{MaT0C21, WuFNL16}. However, the feature selection is not performed obliviously, as it relies solely on hashing function rather than a formal oblivious selection protocol. Subsequently, Yuan et al.~\cite{YuanLQHZX24} employ a pseudorandom function (PRF)~\cite{Pike65a} to realize secure feature selection, and leverage the private data comparison protocol~\cite{ZhengZLGZWSL23} along with tree permutation techniques~\cite{MaT0C21, WuFNL16} to construct a depth-round evaluation scheme.

$\bullet$ \textbf{Other Scenarios:} With the increasing adoption of three-party secure computation~\cite{MohasselR18,BoyleGIN19,FurukawaLNW17,WaghGC19} and function secret sharing (FSS)\cite{GilboaI14,BoyleGI15,BoyleGI16}, many decision tree evaluation protocols have been designed based on three non-colluding cloud servers (O-3PC)\cite{zhang2024secure, JiZLLR23, FuCXSLS24, BaiSZWCCR23, ChengGMMT24}. These protocols can still be broadly categorized into depth-round~\cite{JiZLLR23,FuCXSLS24,BaiSZWCCR23} and constant-round~\cite{JiZLLR23,ChengGMMT24,zhang2024secure} schemes. Leveraging the benefits of replicated secret sharing (RSS), several of these protocols are further enhanced to provide robustness against malicious cloud servers~\cite{FuCXSLS24,BaiSZWCCR23}. Meanwhile, the growing popularity of federated learning~\cite{FLinit,YangLCT19,KonecnyMYRSB16} has sparked significant interest in federated decision tree inference (Fed)~\cite{decisiontree,HEP-XGB,Privet,LuHZWH23,ZhaoZXWLL23}. However, these schemes often overlook the privacy of features or comparisons. For instance, the scheme~\cite{LuHZWH23} only performs plaintext comparisons over decentralized models and requires $2(\tau + 1)$ instances of COT to obtain the oblivious weight. Subsequently, the scheme~\cite{ZhaoZXWLL23} still does not incorporate feature selection. However, it uses the private comparison protocol~\cite{ZhengZLGZWSL23} and performs polynomial-based evaluation~\cite{BostPTG15} using a leveled homomorphic encryption (LHE) scheme~\cite{oldSHE}.

Despite the large number of PDTE schemes that have been proposed, most of them either overlook the challenges of large-scale decision tree evaluation under WAN setting or exhibit poor performance in such scenarios.

\textbf{2) Oblivious Comparison}

As a fundamental building block in decision tree inference, oblivious comparison has been extensively studied and has found broad applications in privacy-preserving machine learning and beyond~\cite{HuangLHD22, Veugen12, Patra0SY21, WaghGC19, ZhengZLGZWSL23, ZhengDWWN22, ZhengWWDN23, LuZS18, Cong00P22, MahdaviNLK23, ShinCLKL24}. Based on their communication complexity, existing comparison protocols can be categorized into three types: $O(t)$-round protocols~\cite{ZhengDWWN22}, $O(\log t)$-round protocols~\cite{ZhengWWDN23, Patra0SY21, HuangLHD22, demmler2015aby}, and $O(1)$-round protocols~\cite{Veugen12, WaghGC19, ZhengZLGZWSL23, LuZS18, Cong00P22, MahdaviNLK23, ShinCLKL24}. Considering the efficiency requirements in WAN setting, we focus on $O(1)$-round protocols. Among them, Wagh et al.~\cite{WaghGC19} propose a protocol for computing the ReLU function that requires five rounds of communication, resulting in notable communication cost. Furthermore, protocols such as~\cite{LuZS18, Cong00P22, MahdaviNLK23, ShinCLKL24} achieve non-interactive comparison by performing computations entirely over ciphertexts. While these non-interactive protocols eliminate communication costs, they suffer from significant inefficiencies due to the computation cost associated with fully homomorphic operations. In contrast, the protocols in~\cite{Veugen12, ZhengZLGZWSL23} require only a single round of interaction. However, the protocol in~\cite{LuZS18} requires bit-wise encryption by DGK, which imposes a heavy bandwidth burden. On the other hand, the protocol proposed by Zheng et al.~\cite{ZhengDWWN22} offers a more balanced trade-off between computation and communication. However, it might be prone to sign errors caused by numerical overflow, and it lacks support for amortized computation, unlike~\cite{Cong00P22, MahdaviNLK23, ShinCLKL24}, thereby limiting its application and performance in large-scale deployments. 

\textbf{3) Secure Path Evaluation}

The current secure path evaluation for decision tree inference mainly contains four types: {Poly}, Path, OnePath, and OnePath$^*$ as~\cref{Summary}. The {Poly} approach requires performing ciphertext multiplications along each path, which demands sufficient multiplicative depth and incurs significant computation cost~\cite{AkaviaLRRSV22,Cong00P22,ZhengDWWN22,BostPTG15,ZhengXWGH23,ZhengWWDN23,ZhaoZXWLL23,TuenoBK20,LiuCLSQ20,LiuSCCSL19}. On the other hand, Path approach necessitate aggregating results across all paths, followed by perturbation and random shuffling to obtain the true weight~\cite{JiZLLR23, KissNLAS19, LuZS18,TaiMZC17,ZhengDWWN22,MahdaviNLK23,zhang2024secure}. Such operations are not well-suited for evaluation over packed ciphertexts (PHE)~\cite{MahdaviNLK23, ShinCLKL24}. In particular, only using a single rotation to shuffle might inadvertently reveal the comparison results of individual paths~\cite{ShinCLKL24}. OnePath$^*$ can reduce the decryption number of Path by enabling single-path evaluation, but its applicability is limited to outsourcing scenarios and relies on a complete binary tree to preserve security~\cite{WuFNL16, ZhaoZWLL23, ChengGMMT24}. While OnePath achieves sublinear computation cost, its performance deteriorates with increasing tree depth, leading to substantial communication cost~\cite{JoyeS18,TuenoKK19,MaT0C21,BaiSCCR22,BaiSZWCCR23,YuanLQHZX24,FuCXSLS24,JiZLLR23}. 

\textbf{4) Support Amortized Computation}

The amortized computation can be categorized into offline amortization and online amortization. In offline computation, the operations are independent of the inputs and can thus be preprocessed to reduce the cost of online computation. For example, in secret-sharing-based multiplication, the generation of multiplication triples can be performed in advance using TTP, OT, or HE~\cite{RatheeR0CGR020,BaiSCCR22,ZhengDWWN22, ZhengWWDN23}. Similarly, in scenarios involving a large number of OT operations, a small number of public-key-based OTs can be used to extend a much larger number of symmetric-key-based OTs, significantly accelerating the online phase~\cite{Pinkas0Z14,CouteauDDKS24,RaghuramanRT23,HuangLHD22}. In online computation, amortization is typically achieved through packed homomorphic encryption (PHE) with single instruction multiple data (SIMD) support, allowing multiple instances to be processed in parallel in a single execution and significantly improving computational efficiency. This approach has been widely adopted in neural network inference~\cite{JuvekarVC18,ZhangXW21,HuangLHD22,BumbleBee}, where it significantly enhances efficiency. Compared to OT-based protocols, PHE reduces communication cost and eliminates the need for any offline phase~\cite{BumbleBee}. However, unlike neural network inference, decision tree evaluation involves strong inter-node dependencies, which limit the effective use of PHE. Consequently, PHE are either underutilized~\cite{LuZS18,AkaviaLRRSV22,ZhaoZXWLL23,ZhaoZWLL23,YuanLQHZX24,TaiMZC17} or not fully exploited~\cite{Cong00P22, MahdaviNLK23, ShinCLKL24} in PDTE.

\section{Preliminaries}
\label{Background}
In this section, we introduce decision tree evaluation, packed homomorphic encryption, private data comparison, and path evaluation. 

\subsection{Decision Tree Evaluation}
Decision tree is an efficient machine learning model and has been widely used in classification or regression tasks~\cite{decision}. A decision tree $\mathcal{T}$ contains three types of nodes: root node, internal nodes, and leaf nodes. Both the root node and the internal nodes are decision nodes. For the $n$-th decision node, it owns the $m[n]$-th feature and a split threshold $y_n$, where $m[n]\in [1,\cdots,M]$ and $M$ is the total number of features. For a leaf node, it owns a leaf weight $w$. A binary tree with $\tau$ decision nodes has $\tau+1$ leaf nodes. Moreover, a full binary tree of depth $D$ has $2^D-1$ decision nodes and $2^D$ leaf nodes. 

Given a feature vector $\mathcal{X} = \{x_1,x_2,\cdots, x_M\}$ with $M$ features, a decision tree evaluation with $\tau$ decision nodes can map the vector into the corresponding leaf weight. The evaluation starts from the root node, and it compares the split threshold $y_{n}$ of the root node ($n=1$) with the feature value $x_{m[n]}$, where $m[n]$ is the feature index of the $n$-th decision node, $m[n] \in [1,\cdots,M]$, and $n\in \{1, \cdots, \tau\}$. Depending on whether the comparison result is 0 ($x_{m[n]} < y_n$) or 1 otherwise, the evaluation direction will go left or right to the next decision node, and continue this comparison until obtaining the evaluation result $\mathcal{T}(\mathcal{X}) \in \{w_1, \cdots, w_{\tau+1}\}$. Through the above rules, a decision path $\mathcal{P}$ with $\mathcal{T}(\mathcal{X})$ can be obtained for the input vector $\mathcal{X}$, and the evaluation is finished. 

\subsection{Packed Homomorphic Encryption}
Packed homomorphic encryption (PHE), such as BFV~\cite{BFV}, BGV~\cite{BGV}, CKKS~\cite{CKKS}, etc, is a homomorphic encryption technology that supports batch processing of multiple data. In this paper, we focus on BFV cryptosystem~\cite{BFV}. It consists of three algorithms, including Key Generation, Encryption, Decryption, and it operates over plaintext space $\mathcal{R}_q = \mathcal{Z}_q[x]/(x^\mathtt{N}+1)$ and ciphertext space $\mathcal{R}_Q = \mathcal{Z}_Q[x]/(x^\mathtt{N}+1)$, where $q$ is the plaintext modulo and $Q$ is the ciphertext modulo. Moreover, there exist several distributions in BFV as follows, $\textbf{1)}$ $\mathcal{D}_1$ is a key distribution; $\textbf{2)}$ $\mathcal{D}_2$ is an error distribution; $\textbf{3)}$ $\mathcal{D}_Q$ is a uniform random distribution over the ciphertext space $\mathcal{R}_Q$. 

$\bullet$ $\mathtt{KeyGen(\mathtt{\lambda}, \mathtt{N})}.$ Given a security parameter $\mathtt{\lambda}$ and a polynomial size $\mathtt{N}$, generate the distributions $\{\mathcal{D}_1, \mathcal{D}_2,\mathcal{D}_Q\}$, the private key $\mathtt{s\leftarrow \mathcal{D}_1}$ and the public key $\mathtt{pk = (b,a) = (as+e, a)}$, where $\mathtt{e} \leftarrow \mathcal{D}_2$ and $\mathtt{a}\leftarrow \mathcal{D}_Q$.

$\bullet$ $\mathtt{Enc(m, pk)}.$ Given a plaintext $\mathtt{m}\in \mathcal{Z}_q^\mathtt{N}$, the algorithm encodes $\mathtt{m}$ into a polynomial $\textbf{m}$ with SIMD~\cite{SIMD}. It samples a random polynomial $\mathtt{r} \leftarrow \mathcal{D}_1$ and two noise polynomials $\mathtt{e}^{(1)}, \mathtt{e}^{(2)} \leftarrow \mathcal{D}_2$. Then, it encrypts $\textbf{m}$ into the ciphertext $\mathtt{c} = \llbracket \mathtt{m} \rrbracket = \mathtt{(c_0,c_1)} = \mathtt{(br+\textbf{m}+e^{(1)}, ar+e^{(2)})}$. Finally, it returns the ciphertext $\mathtt{c}$.

$\bullet$ $\mathtt{Dec(c, s)}.$ The polynomial $\textbf{m}$ can be recovered as $\textbf{m} = \mathtt{c_0 + c_1s}$. Finally, it returns the decoded plaintext $\mathtt{m}$.

We utilize $\llbracket \mathtt{m_1} \rrbracket + \textbf{m}_2$, $ \llbracket \mathtt{m_1} \rrbracket\circ \textbf{m}_2$, $\llbracket \mathtt{m_1} \rrbracket + \llbracket \mathtt{m_2} \rrbracket$, and $\llbracket \mathtt{m_1} \rrbracket \circ \llbracket \mathtt{m_2} \rrbracket$ to represent plaintext-ciphertext addition, plaintext-ciphertext multiplication, ciphertext addition, and ciphertext multiplication, where $\llbracket \mathtt{m_1} \rrbracket,\llbracket \mathtt{m_2} \rrbracket$ are two BFV ciphertexts, $\textbf{m}_2$ is an encoded plaintext vector, and $\mathtt{m_1}, \mathtt{m_2}\in \mathcal{Z}_q^\mathtt{N}$. We utilize $\mathtt{Rot(c}, r)$ to represent the ciphertext rotation, where the ciphertext slots are shifted by a random integer $r$. We utilize $\mathtt{plainRot(m}, r)$ to represent the plaintext rotation. When $r>0$, the slots are shifted to the right, otherwise left. For simplicity, encoded vectors are denoted using bold uppercase symbols.

\subsection{Private Data Comparison}
\label{comparisonPrivate}
The private data comparison has recently gained popularity~\cite{ZhengZLGZWSL23,ZhaoZWLL23,ZhaoZXWLL23}. It enables one party $S_1$ with $\{\llbracket \mathtt{m_1} \rrbracket, \llbracket \mathtt{m_2} \rrbracket\}$ and the other party $S_2$ with the secret key $s$ to compare $\mathtt{m_1}$ and $\mathtt{m_2}$. Finally, $S_1$ obtains the comparison result. It consists of two steps.

$\textbf{Step 1.}$ $S_1$ calculates $\llbracket \mathtt{m} \rrbracket = a\cdot (\llbracket \mathtt{m_1} \rrbracket - \llbracket \mathtt{m_2} \rrbracket) + b$ and sends $\llbracket \mathtt{m} \rrbracket$ to $S_2$, where $\{a,b\}$ are selected from the plaintext space, with $a>b>0$.

$\textbf{Step 2.}$ After receiving $\llbracket \mathtt{m} \rrbracket$, $S_2$ calculates the comparison result $c$ and sends it to $S_1$. Here, $c = 0$ if $\mathtt{m} < 0$, indicating that $\mathtt{m_1} < \mathtt{m_2}$, and $c=1$ if $\mathtt{m} > 0$, indicating that $\mathtt{m_1} \geq \mathtt{m_2}$.

\subsection{Path Evaluation}
In this section, we introduce the path cost-based evaluation (Path)~\cite{TaiMZC17} and polynomial-based evaluation ({Poly})~\cite{BostPTG15} in details. The concept of path cost-based evaluation is as follows: given a decision tree model $\mathcal{T}$ and a feature vector $\mathcal{X}$, we determine the comparison result $c_n$ (is $x_{m[n]} < y_{n}$ ? $0 : 1$) for each decision node. For the $n$-th decision node, the $\mathtt{node}$\texttt{->}$\mathtt{left.cost}$ is set to $c_n$ and the $\mathtt{node}$\texttt{->}$\mathtt{right.cost}$ is set to $1 - c_n$. As the cost of each node is assigned a value, the cost from the root node to a corresponding leaf node is defined as the sum of the costs along the path. It can be observed that only the sum of the true path equals zero, whereas the sums of all other paths are greater than zero. Thus, the weight of the leaf node along this path corresponds to the predicted weight. Unlike path cost-based evaluation, the $n$-th $\mathtt{node}$\texttt{->}$\mathtt{left.cost}$ is set to $1 - c_n$ and the $n$-th $\mathtt{node}$\texttt{->}$\mathtt{right.cost}$ is set to $c_n$. Moreover, the cost from the root node to a corresponding leaf node is defined as the product of the costs along the path. It can be observed that only the product of the true path equals one, while the products of all other paths are zero. 

\section{Framework Overview}
In this section, we begin by formulating the key challenges addressed by Kangaroo. We then present a high-level overview of its workflow, followed by a detailed description of our models and corresponding security analysis. \cref{Notations} presents the corresponding notations used in Kangaroo.

\subsection{Technique Challenges and Observations}
\label{Motivation}
While prior studies have explored PHE-based model inference~\cite{LuZS18, Cong00P22, MahdaviNLK23, AkaviaLRRSV22, ShinCLKL24}, many of these solutions fail to fully exploit the potential of SIMD techniques~\cite{LuZS18,Cong00P22,MahdaviNLK23,AkaviaLRRSV22} or suffer from low packing utilization and limited execution efficiency~\cite{Cong00P22,MahdaviNLK23,ShinCLKL24}. For example, the schemes~\cite{LuZS18, Cong00P22, MahdaviNLK23} cannot support the amortized calculation because XCMP cannot support SIMD. t-SortingHat~\cite{Cong00P22} supports only the comparison amortization, but its complexity remains high at $O(\frac{\tau \cdot t}{\log \tau})$. RCC-PDTE~\cite{MahdaviNLK23} and the scheme proposed in~\cite{ShinCLKL24} support amortization for both comparison and path evaluation. However, due to limitations in numerical precision and path length, these schemes fail to fully leverage the parallelism available in each ciphertext slot. Therefore, we identify two core challenges: amortized utilization and computational efficiency. To enhance the amortization capability of PHE, we should treat each ciphertext slot as a representation of a decision tree node and avoid redundant node representations. Consequently, node information must be encoded into the coefficients of the ciphertexts, enabling efficient feature selection and secure comparison directly over packed ciphertexts. However, even with successful packing, efficient path evaluation over packed ciphertext remains a non-trivial problem, leading to many existing schemes to favor Paillier (AHE)~\cite{ZhaoZWLL23,TaiMZC17,YuanLQHZX24} over PHE for constructing constant-round evaluation schemes. In what follows, we elaborate on the technical challenges and observations.

$\bullet$ \textbf{Challenge and Observation 1:} When using HE to encrypt each feature eliminates the need for feature selection and enables direct comparison. Therefore, using PHE introduces the first challenge: how to perform parallelized and secure feature selection within packed ciphertexts and efficiently complete the comparison. To tackle this, we quantize the feature vectors and redundantly encode them into polynomials. Furthermore, we design a non-interactive feature selection algorithm to enable batch processing of feature selection across nodes. It supports feature selection across arbitrary dimensions and places the results at fixed positions to hide the features of the nodes. To further enhance the amortization capability of PHE, we encode the nodes of multiple trees into a single polynomial and simultaneously pack multiple selected feature vectors into a single polynomial. This design allows us to treat each coefficient in the polynomial as a unique, non-redundant node, thereby fully leveraging the SIMD parallelism and amortization potential offered by PHE. Based on the packed selected features and thresholds, we integrate secret sharing and propose a packed oblivious comparison protocol to achieve correct and secure comparisons, which solve the limitation of the scheme~\cite{ZhengZLGZWSL23} mentioned in~\cref{relatedwork}. Moreover, our protocol performs comparisons using SIMD in a single round of interaction, which further amortizes both computation and communication costs.

$\bullet$ \textbf{Challenge and Observation 2:} After obtaining the encrypted comparison results, the server typically performs either path cost-based evaluation or polynomial-based evaluation to determine the final path selection. For path cost-based evaluation over packed ciphertexts, it is necessary to sum the evaluation results across all nodes along each path, which requires a significant number of rotation and summation operations. Additionally, to conceal the final selected path, the selected weight must undergo extensive rotations, leading to an impractically high computation cost. In contrast, polynomial-based evaluation can generate a one-hot vector to get the selected weight, eliminating the aforementioned costs. However, it requires a multiplication depth corresponding to the tree depth for each path, and its computation cost cannot be amortized. As a result, both methods impose significant computation cost. To address this challenge, we propose a novel path evaluation protocol that enables the client and server to perform plaintext evaluations on an obfuscated tree model, yielding results similar to those of path cost-based evaluation. Simultaneously, we employ the oblivious comparison protocol to convert the evaluation result into a polynomial representation, thereby avoiding excessive rotations. 

\begin{figure} 
    \centering
    \includegraphics[width=\linewidth]{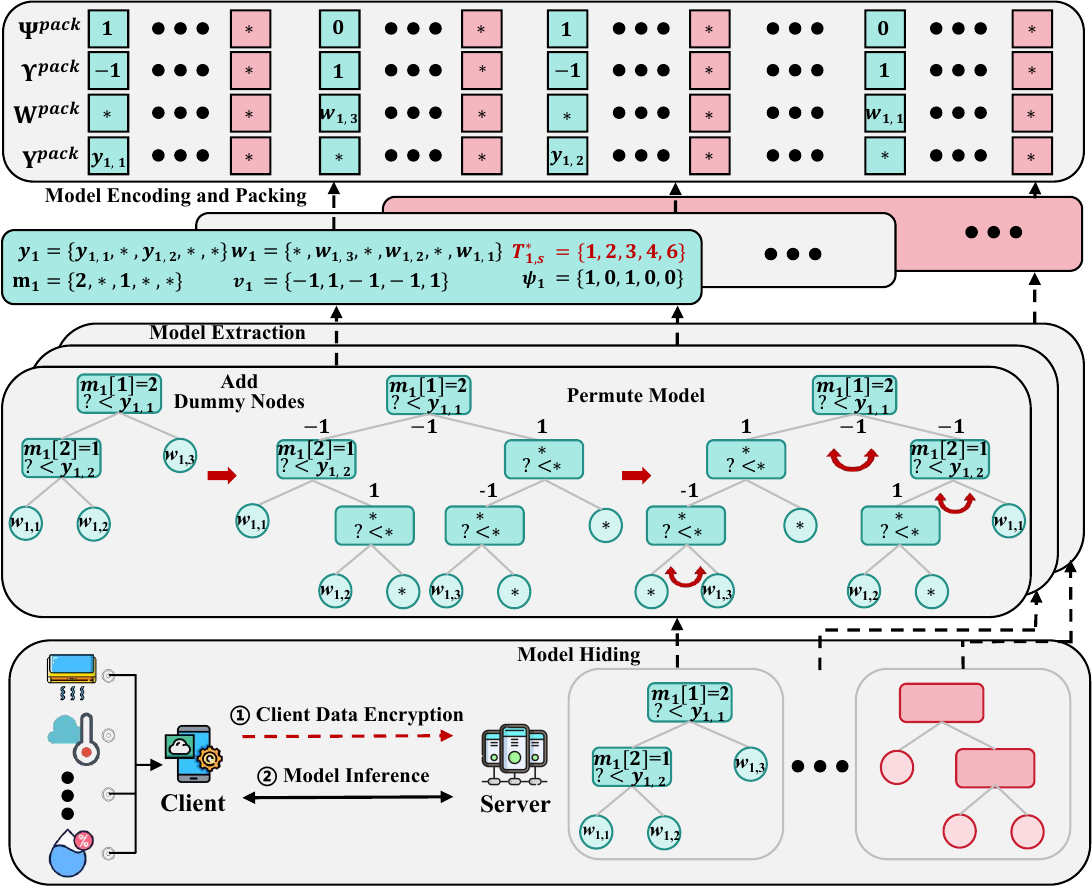}
    \caption{High-level workflow of Kangaroo for client-server model, where $K= M$.}
    \label{systemModel}
\end{figure}

\subsection{Kangaroo Workflow}
\label{workflow}
The processing flow of Kangaroo consists of four stages, model hiding and extraction, model encoding and packing, client data encryption, and model inference. Below, we provide a concise overview of the main operations involved in each phase, which are also depicted in~\cref{systemModel}.

$\bullet$ \textbf{Model Hiding and Extraction:} To finish the path evaluation, the server needs to share the tree model structure with the client. Directly sharing the model structure might expose the server's model privacy~\cite{LiuJLA17,JuvekarVC18,ZhangXW21}. Expanding decision nodes into a full binary tree~\cite{WuFNL16,JoyeS18,ZhengDWWN22,ZhengWWDN23,ZhaoZWLL23} offers a potential solution but introduces significant computation costs. To address it, we introduce a more efficient solution to hide the model structure. It avoids the need to expand the nodes into a full binary tree, and instead only requires publishing an obfuscated model structure to effectively protect the original model structure. Specifically, the server needs to generate some dummy nodes to hide the model structure. We assume that the total number of decision nodes, after padding with dummy nodes, is $\tau^*$. When a dummy node is created, its left leaf inherits the weight of the parent node, while its right leaf's weight is randomly generated. Then, the server confuses the tree model by a coin-flipping method. For each node, the server flips a coin to randomly select either $1$ or $-1$. If $-1$ is selected, the server swaps the left and right child nodes of the given node; otherwise, no change is made. 

During model extraction, the server extracts the model parameters of the $k$-th tree, where $1\leq k\leq K$. Specifically, the server uses breadth-first to obtain the threshold $\textbf{y}_k$, feature index $\textbf{m}_k$, flip condition $\upsilon_k$, model structure index $\mathcal{T}^*_{k,s}$, and status of node $\psi_k$ and uses depth-first traversal to obtain the weight vector $\textbf{w}_k$ of the leaf nodes for $k$-th tree. In the vector $\psi_k$, $0$ indicates a dummy node and $1$ indicates a real node. The model structure index $\mathcal{T}^*_{k,s}$ is obtained according to the following rule: for a node with index $i$, its left child index is $2i$, and its right child index is $2i+1$. At last, the model structure indices $\{\mathcal{T}^*_{k,s}\}_{k = 1}^K$ are published. 

$\bullet$ \textbf{Model Encoding and Packing:} To amortize the time cost of feature selection, comparison, and path evaluation, the server encodes the model parameters into a polynomial ring. {First, the server chooses the corresponding the maximum and minimum quantization vectors $\mathcal{X}^{max} = \{x_1^{max},x_2^{max},\cdots, x_M^{max}\}$ and $\mathcal{X}^{min} = \{x_1^{min},x_2^{min},\cdots, x_M^{min}\}$ and the precision parameter $\zeta$. Then, each vector $\textbf{y}_k$ is quantized by $\mathcal{X}^{max}$ and $\mathcal{X}^{min} $ with the precision parameter $\zeta$ to enable secure comparison. It is worth noting that the server needs to publish $\mathcal{X}^{max}$, $\mathcal{X}^{min}$, and $\zeta$. For non-sensitive features, such as human height, the valid range (e.g., 54.6 cm to 272 cm) can be directly published without privacy concerns. However, for certain sensitive features, the actual range can be deliberately extended to obscure the true value range and enhance privacy protection.} Specifically, $\textbf{y}_k$ is quantized as
\begin{equation*}
    \textbf{y}_{k} = \{\frac{y_{k}[1] - x_{m_{k}[1]}^{min}}{x_{m_{k}[1]}^{max} - x_{m_{k}[1]}^{min}}\cdot \zeta, \cdots, \frac{y_{k}[\tau^*] - x_{m_{k}[\tau^*]}^{min}}{x_{m_{k}[\tau^*]}^{max} - x_{m_{k}[\tau^*]}^{min}}\cdot \zeta\}.
\end{equation*}
After it, the vectors $\textbf{y}_k$, $\upsilon_k$, and $\psi_k$ are encoded into $\{\textbf{Y}_k, \Upsilon_k, \Psi_k\}$ for $1\leq k\leq K$, where each element of the original vectors is placed at positions $\{(n-1)M+1\}_{n=1}^{\tau^*}$. The vector $\textbf{w}_k$ is encoded into $\textbf{W}_k$ for $1\leq k\leq K$, with elements placed at positions $\{(n-1)M+1\}_{n=1}^{\tau^*+1}$. The vector $\textbf{m}_k$ is encoded into $\textbf{M}_k = \{\textbf{m}^1_k, \textbf{m}^2_k, \cdots, \textbf{m}^{\tau^*}_k\}$, where $\textbf{m}^n_k$ is a vector of $M$ dimensions and $1\leq n\leq \tau^*$. The $m_{k}[n]$-th value of $\textbf{m}^n_k$ is set to $1$, and other values are set to $0$.

To further improve the utilization of the polynomial space, the server packs multiple tree models together for efficient inference. Specifically, the server merges the vectors as $\textbf{Y}_{\gamma}^{pack} = \sum_{m=1}^M \mathtt{plainRot} (\textbf{Y}_{(\gamma-1)M+m}, m-1)$, $\textbf{W}_{\gamma}^{pack} = \sum_{m=1}^M \mathtt{plainRot} (\textbf{W}_{(\gamma-1)M+m}, m-1)$, $\Upsilon_{\gamma}^{pack} = \sum_{m=1}^M \mathtt{plainRot} (\Upsilon_{(\gamma-1)M+m}, m-1)$, and $\Psi_{\gamma}^{pack} = \sum_{m=1}^M \mathtt{plainRot} (\Psi_{(\gamma-1)M+m}, m-1)$. For simplicity, we assume $K = \Gamma \cdot M$ and refer to $\{\textbf{Y}_{\gamma}^{pack}, \textbf{W}_{\gamma}^{pack}, \Upsilon_{\gamma}^{pack}, \Psi_{\gamma}^{pack}\}$ as the packed model, where $1\leq \gamma \leq \Gamma$. After getting the packed model and $\{\textbf{M}_k\}_{k=1}^K$, a same-sharing-for-same-model strategy is employed to improve the efficiency of evaluation, which is introduced in \cref{samesharing}.

$\bullet$ \textbf{Client Data Encryption:} During the data encryption, the client needs to initialize a BFV encryption system and generates the public key to request evaluation services. Then, the client utilizes the public range of the feature vector $\mathcal{X}^{max}$ and $\mathcal{X}^{min}$ to quantize its feature vector $\mathcal{X}$ as 
\begin{equation*}
    \mathcal{X} = \{\frac{x_1 - x_1^{min}}{x_1^{max} - x_1^{min}}\cdot \zeta, \cdots,\frac{x_M - x_M^{min}}{x_M^{max} - x_M^{min}} \cdot \zeta\}.
\end{equation*}
After it, the client repeats the quantized feature vector $\tau^*$ times to an $\mathtt{N}$-dimensional vector $X$, where $X= \{\mathcal{X}, \cdots, \mathcal{X}\}$. We assume $(\tau^* +1)\cdot M \leq \mathtt{N}$, and $\mathtt{N}$ is the polynomial dimension of BFV. At last, the client encrypts the vector and sends $\llbracket X \rrbracket$ and the corresponding keys to the server.

$\bullet$ \textbf{Model Inference:} During the model inference, the server and the client jointly execute feature selection, comparison, and path evaluation, and the client obtains the inference result. The building blocks involved in the inference contain three steps: \ding{172} $\mathtt{PackFeatureSel}$, \ding{173} $\mathtt{PackObliviousCom}$, and \ding{174} $\mathtt{PackPathEva}$. They serve as the core operations for performing inference on a single decision tree. $\mathtt{PackFeatureSel}$ is used to select the corresponding feature over the encrypted and packed feature vector, with the selected values placed in fixed positions to hide the node's feature information. $\mathtt{PackObliviousCom}$ then performs packed comparisons between feature thresholds and selected features for each node and obtains the encrypted comparison signs. In the signs, $0$ indicates that the client's feature is less than the threshold, and 1 otherwise. Finally, $\mathtt{PackPathEva}$ is applied to obtain the evaluation result based on the encrypted comparison outcomes. The weight corresponding to the true decision path is retrieved, while the weights at all other positions are set to $0$. Based on the building blocks, a complete scheme for large-scale evaluation is proposed, along with several optimization techniques to accelerate the process. Moreover, some additional functionalities are presented in the appendix, and interested readers are encouraged to refer to it for more details.

\subsection{Models and Security}
\label{Modelandsecurity}
The Kangaroo mainly works in the client-server model. The server holds $K$ decision tree models $\{\mathcal{T}_k\}_{k=1}^K$, and the client holds a feature vector $\mathcal{X}$. Our security goals are to protect the server's models from the semi-honest client and the client's data from the semi-honest server. We consider semi-honest adversaries, which are similar to previous works~\cite{BostPTG15,TaiMZC17,LuZS18,KissNLAS19,TuenoKK19,Cong00P22,MahdaviNLK23}. Specifically, the following private data should be protected.

\textbf{1) The Private Data of Server}

\ding{172} $\textbf{m}_k:$ The node feature vector is private for server as it reflects the splitting preferences of the decision tree and might expose important feature distributions.

\ding{173} $\textbf{y}_{k}:$ The feature threshold vector is private for server as it determines how data points are classified, and leakage could expose critical decision boundaries.

\ding{174} $\textbf{w}_k:$ The weight vector is private for server as it  encapsulates critical information about the model.

\ding{175} $\mathcal{T}_{k,s}, \upsilon_k, \psi_k:$ The model structure index, flip condition, and node status are private for server as they represent the server's revenue model, and revealing them could leak some sensitive information about the private dataset used to train the tree model~\cite{LiuJLA17,JuvekarVC18,ZhangXW21}.
 
\textbf{2) The Private Data of Client}

\ding{176} $\mathcal{X}:$ The feature vector is private for client as it contains the client's private input data used for inference.

\textbf{3) The Private Data during Inference}

\ding{177} The evaluation result is private for client as it reveals the models' decision on the client's private feature vector.

\ding{178} The decision path is sensitive for server and client as it reveals how the client's input traverses each decision tree and exposes the corresponding weights to the client.

\ding{179} The comparison sign is sensitive for server and client as it might reveal the decision direction.

\ding{180} The difference of comparison is sensitive for server and client as it directly involves private feature value and its relation to the decision threshold.

\textbf{Security of Kangaroo:} {To protect the above privacy, the private data of client should be encrypted and the private data of server should be blinded~\cite{TaiMZC17,LiuJLA17,KissNLAS19}. We present the standard definition of semi-honest security and conduct a rigorous analysis using the simulation-based real/ideal world model~\cite{Foundations}. In this setting, the client and server are assumed to strictly follow the protocol but may attempt to infer sensitive information as defined in our threat model. In real-world scenarios, however, a curious client may craft inputs in a malicious way to extract model parameters. To mitigate such risks, it is essential that all intermediate outputs in our protocol remain oblivious~\cite{araki2016high,WaghGC19}. In either of the above cases, our scheme should be designed to robustly protect private data \ding{172} - \ding{180}.} Due to space limitation, the details can be found in~\cref{secureDefinition,secure1}. 

Kangaroo also supports the outsourcing scenarios, which operates in the single-cloud assisted model~\cite{ZhaoZWLL23}. The server outsources the model to a cloud service provider (CSP) to offload client-side computation and reduce its own computational burden. It can also be extended to a double-cloud model~\cite{MaT0C21,ZhengDWWN22,ZhengWWDN23}. However, considering the cost of leasing servers and the security issues associated with the double-cloud model, we have chosen the single-cloud assisted outsourcing scheme. The interested readers can refer to~\cref{outsourceKangaroo,secure2} for more details.

\section{Building Blocks}
\label{buildblocks}
In this section, we introduce our building blocks for single tree inference, including packed feature selection $\mathtt{PackFeatureSel}$, packed oblivious comparison $\mathtt{PackObliviousCom}$, and packed path evaluation $\mathtt{PackPathEva}$, to achieve the amortized computation.

\subsection{Packed Feature Selection: $\mathtt{PackFeatureSel}$}
\begin{algorithm}[!t]
	\renewcommand{\algorithmicrequire}{\textbf{SInput:}}
	\renewcommand{\algorithmicensure}{\textbf{SOutput:}}
	\caption{$\mathtt{I}$-$\mathtt{PackFeatureSel}$}
	\label{PackFeatureSel}
	\begin{algorithmic}[1]
        \Require The encrypted vector $\llbracket X \rrbracket$, feature size $M$, and encoded feature index vector $\textbf{M}$.
		\Ensure The selected encrypted vector $\llbracket X' \rrbracket$.
        \LComment{The server executes:}
        \State $\llbracket X' \rrbracket = \llbracket X \rrbracket \circ  \textbf{M} $.
        \State $bool = (M > 0$ $\&$ $(M $ \& $(M-1)) == 0)$. \Comment{Determine whether M is a power of 2.}
        \If{$bool == true$}
            \For{$i = 0, i < \log M, i ++$}
                \State $\llbracket X' \rrbracket = \llbracket X' \rrbracket + \mathtt{Rot}(\llbracket X' \rrbracket, -2^{i})$.
            \EndFor
        \Else
            \State $\llbracket X'' \rrbracket = \llbracket X' \rrbracket + \mathtt{Rot}(\llbracket X' \rrbracket, -1)$.
            \If{$M\mod 2 == 0$}
                \State $\llbracket X' \rrbracket = \llbracket X'' \rrbracket$.
            \EndIf
            \For{$i = 1, i < \lceil \log M \rceil - 1, i ++$}
                \State $M = M - \lfloor \frac{M}{2} \rfloor$.
                \If{$M\mod 2 == 0$}
                    \State $\llbracket X' \rrbracket = \llbracket X'' \rrbracket + \mathtt{Rot}(\llbracket X' \rrbracket, -2^{i})$.
                \EndIf
                \State $\llbracket X'' \rrbracket = \llbracket X'' \rrbracket + \mathtt{Rot}(\llbracket X'' \rrbracket, -2^{i})$.
            \EndFor
            \State $\llbracket X' \rrbracket = \llbracket X'' \rrbracket + \mathtt{Rot}(\llbracket X' \rrbracket, -2^{\lceil \log M \rceil-1})$.
        \EndIf
        \State The server gets $\llbracket X' \rrbracket$.
	\end{algorithmic}   
\end{algorithm}
The $\mathtt{PackFeatureSel}$ is used to select the corresponding features from the client's packed feature vector. The server inputs an encrypted vector $\llbracket X \rrbracket$, a feature size $M$, and an encoded feature index vector $\textbf{M}$, and then obtains a selected encrypted vector $\llbracket X' \rrbracket$. In $\llbracket X' \rrbracket$, the selected features $x_{m[n]}$ are placed at positions $\{(n-1)M+1\}_{n=1}^{\tau^*}$ to hide the features of the nodes. We give a non-interactive feature selection algorithm $\mathtt{I}$-$\mathtt{PackFeatureSel}$ in \cref{PackFeatureSel}. First, the server computes $\llbracket X' \rrbracket = \llbracket X \rrbracket \circ  \textbf{M} $ to extract the corresponding feature values from the client at each node as line $2$. To accumulate each vector of length $M$ to its first value, the server checks whether $M$ is a power of $2$ as line $3$. If yes, the server performs $\log M$ rotate-and-sum operations on $\llbracket X' \rrbracket$ to place the extracted results at positions $\{(n-1)M+1\}_{n=1}^{\tau^*}$ as lines $4-6$. If no, a division-based approach is applied to compute the selected encrypted vector as lines $7-16$. In the process, the server sums every two adjacent data. Therefore, it needs to determine whether $M$ is odd or even for the vector of length $M$. If $M$ is odd, the value at the last position will be retained in $\llbracket X' \rrbracket$. If $M$ is even, every two adjacent data will be summed. Using the division approach, the server can obtain the sums of the first $\frac{M}{2}$ values into $\llbracket X'' \rrbracket$ and stores the remaining sums in $\llbracket X' \rrbracket$. Finally, the server rotates $\llbracket X' \rrbracket$ and adds it to $\llbracket X'' \rrbracket$, obtaining the final result. The correctness of $\mathtt{PackFeatureSel}$ is proven in \cref{theoremOne}. Moreover, we present a fully amortized feature selection as~\cref{PackFeatureSel1}. 

\begin{figure}[t]
    \begin{tcolorbox}[colback = white, colframe = lightgray]
        \textbf{$\mathtt{PackObliviousCom}$ Protocol} \\ 
        \textbf{SInput:} The encrypted and encoded vectors $\llbracket X' \rrbracket$, $\textbf{Y}$. \\
		\textbf{SOutput:} The encrypted comparison result vector $\llbracket C \rrbracket$. \\
        $\color{gray}\rhd$ $\color{gray}\textit{The server executes:}$ \\
        {\small1:} $A \leftarrow \{a_{1},*,\cdots,*, \cdots,a_{(\tau^*-1)M+1},*,\cdots,*\}$, $B\leftarrow \{b_{1},*,$ $\cdots,*, \cdots,b_{(\tau^*-1)M+1},*,\cdots,*\}$, where $\zeta > A[{(n-1)M+1}] > B[{(n-1)M+1}] > 0$ and $1\leq n\leq \tau^*$. \Comment{Random the differences results.} \\
        {\small2:} $R\leftarrow \{r_{1},*,\cdots,*, \cdots,r_{(\tau^*-1)M+1},*,\cdots,*\}$, where $R[{(n-1)M+1}] \leftarrow \{1,-1\}$ by flipping a coin. \Comment{Random the signs of comparison results.}  \\
        {\small3:} $\llbracket V \rrbracket \leftarrow  \textbf{A}\circ \textbf{R} \circ (\llbracket X' \rrbracket - \textbf{Y}) + \textbf{B}\circ \textbf{R} \Rightarrow$ the client. \\
        $\color{gray}\rhd$ $\color{gray}\textit{The client executes:}$ \\
        {\small4:} $V \leftarrow \mathtt{Dec}(\llbracket V \rrbracket, \mathtt{s}) $, $V' \leftarrow \{v'_{1},*,\cdots,*, \cdots, \\v'_{(\tau^*-1)M+1} ,*,\cdots,$ $*\}$, where $V'[{(n-1)M+1}] = 0$ if $V[(n-1)M+1] < 0$, $1$ otherwise. \\
        {\small5:} $\llbracket V'\rrbracket \leftarrow \mathtt{Enc}(V', \mathtt{pk} ) \Rightarrow$ the server. \\
        $\color{gray}\rhd$ $\color{gray}\textit{The server executes:}$ \\
        {\small6:} $C' \leftarrow \{c'_{1},*,\cdots,*, \cdots,c'_{(\tau^*-1)M+1},*,\cdots,*\}$, where $C'[{(n-1)M+1}] = 1$ if $R[{(n-1)M+1}] = -1$, $0$ otherwise. \\
        {\small7:} The server gets $\llbracket C \rrbracket \leftarrow \textbf{C}' +  \textbf{R}  \circ \llbracket V' \rrbracket$.
    \end{tcolorbox}
    \caption{Packed oblivious comparison protocol for single tree.}
    \label{PackObliviousCom}
\end{figure}

\subsection{Packed Oblivious Comparison: $\mathtt{PackObliviousCom}$}
The $\mathtt{PackObliviousCom}$ is used to achieve the oblivious comparison between the node thresholds and client's selected features. The server inputs $\llbracket X' \rrbracket$ and an encoded vector $\textbf{Y}$, and then obtains the encrypted comparison result vector $\llbracket C \rrbracket$, where $0\leq X'[(n-1)M+1] = x_{m[n]}, Y[(n-1)M+1] = y_n \leq \zeta$. In $\llbracket C \rrbracket$, $C[(n-1)M+1] = 0$ if $x_{m[n]}-y_n < 0$, $1$ otherwise, where $1\leq n \leq \tau^*$. The $\mathtt{PackObliviousCom}$ is given in \cref{PackObliviousCom}. First, the server selects two random vectors $A$ and $B$ to blind the difference result of each node as line $1$. Then, the server randomly selects either $1$ or $-1$ for each node by flipping a coin to protect the result's sign as line $2$. Next, the server blinds the difference result $\llbracket X' \rrbracket - \textbf{Y}$ and gets $\llbracket V \rrbracket$ as line $3$. The $\llbracket V \rrbracket$ is sent to the client. On receiving the $\llbracket V \rrbracket$, the client decrypts it and obtains the blinded comparison result at each node. The client sets $v'_{n}$ to $0$ if $V[(n-1)M+1] < 0$, $1$ otherwise, to get the vector $V'$ as line $4$. Then, the client encrypts $V'$ and returns it to the server as line $5$. After obtaining $\llbracket V' \rrbracket$, the server generates the vector $C'$ as line $6$. At last, the server utilizes $C'$ and $R$ to recover the comparison result vector $\llbracket C \rrbracket$. The correctness of $\mathtt{PackObliviousCom}$ is proven in \cref{theoremTwo}. 

$\textit{Remark 1.}$ It is obvious that our $\mathtt{PackObliviousCom}$ can also support comparison of two ciphertexts. In addition, we can attach specific values to $*$ for more amortization.

\begin{table*}
    \caption{Online operation costs for two-party inference schemes: $\tau:$ the number of decision nodes; $M:$ the feature dimension; $t:$ the bit size of feature; $K:$ the number of trees; $D:$ the tree depth; $\mathtt{Mul}^*:$ plaintext-ciphertext multiplication. SOS: Shard Oblivious Selection Protocol~\cite{BaiSCCR22}; DGK: Private Comparison Protocol~\cite{DGK}.}
    \label{Complexity}
    \centering
    \small
    \begin{threeparttable}
    \footnotesize
    \begin{tblr}
        {
            width=\textwidth,
            colspec         = {*{5}{Q[c,m]}Q[c,m]},
            row{1,Z}      = {font=\bfseries},
            hline{2}        = {1-Z}{wd=.1em,leftpos=1,rightpos=1,endpos=true},
        }
        \hline                     
                                              & Primitives  & Selection                                       & Comparison                            &   Path evaluation        &    Round                           \\
        Ma~\cite{MaT0C21} (Sparse)            & SS,GC,OT    & $KD\cdot$(SS, $\binom{M}{1}$-OT)                & $K D\cdot$ (GC, $\binom{2}{1}$-OT)    &     /                    & 2D-1      \\
        Bai~\cite{BaiSCCR22} (HE-SOS)         & SS,OT,AHE   & $KD\cdot$ SOS                                   &  $KD\cdot$ (GC, SOS)                  &   /                      & 8D      \\
        \hline
        Cong~\cite{Cong00P22} (Sortinghat)    & PHE         & /                                               & $K \tau\cdot\mathtt{Mul}^*$           & $O(K\tau) (\mathtt{Mul} + \mathtt{Add})$          & 1     \\
        Mahdavi~\cite{MahdaviNLK23} (Levelup) & PHE         & /                                               & $K \tau\cdot \mathtt{Mul}^*$          & $O(K\tau) (\mathtt{Add} + \mathtt{Mul}^*)$             & 1     \\
        Tai~\cite{TaiMZC17} (HHH)             & AHE         & /                                               & $K \tau\cdot$ DGK                     & $O(K\tau) (\mathtt{Add} + \mathtt{Mul}^*)$     & 2   \\
        Kiss~\cite{KissNLAS19} (GGH)          & GC,AHE,OT   & $KMt\cdot($GC + $\binom{2}{1}$-OT$)$                                                    & $K \tau\cdot$ GC                    & $O(K\tau) (\mathtt{Add} + \mathtt{Mul}^*)$           & 2     \\
        Kiss~\cite{KissNLAS19} (HGH)          & GC,SS,AHE   & $KM\tau\cdot(\mathtt{Add}+\mathtt{Dec})$        & $K \tau\cdot$ GC                      & $O(K\tau) (\mathtt{Add} + \mathtt{Mul}^*)$             & 3      \\
        \hline
        Ours (Kangaroo)                & PHE,SS       & $2K\cdot\mathtt{Mul}^*+ O(K\log M) \mathtt{Rot}$ & $\lceil \frac{K}{M} \rceil(2\cdot \mathtt{Mul}^* + \mathtt{Enc}+ \mathtt{Dec})$ & $2\lceil \frac{K}{M}\rceil (\mathtt{Mul}^* + \mathtt{Enc}+ \mathtt{Dec})$               & 4         \\
        \hline
    \end{tblr}
    \begin{tablenotes}
        \footnotesize
        \item[1] In the schemes~\cite{Cong00P22,MahdaviNLK23}, we only analyze the operation costs of XCMP.
    \end{tablenotes}
    \end{threeparttable}
\end{table*}

\subsection{Packed Path Evaluation: $\mathtt{PackPathEva}$}
The $\mathtt{PackPathEva}$ is used to get the evaluation result. The server inputs $\llbracket C \rrbracket$, an structure index $\mathcal{T}_s^*$, and an encoded weight vector $\textbf{W}$, and the client inputs $\mathcal{T}_s^*$. After executing it, the server obtains the encrypted evaluation result vector $\llbracket T \rrbracket$. In $T$, only one element is the actual weight value, and the remaining elements are $0$. {Due to space limitation, the $\mathtt{PackPathEva}$ is given in~\cref{PackPath}.} First, the server selects a random vector $R'$ to blind the comparison result vector $\llbracket C \rrbracket$. Then, the server sends the blinded result $\llbracket I' \rrbracket$ to the client as line $1$. The client constructs a tree by $\mathcal{T}_s^*$. After receiving $\llbracket I' \rrbracket$, the client decrypts and assigns the corresponding left cost and right cost for the $n$-th decision node as line $2$. The client sums costs along each path to obtain $I''$, encrypts and sends $\llbracket I'' \rrbracket$ to the server as lines $3-4$. Similarly, the server also constructs a tree by $\mathcal{T}_s^*$ and sums the perturbation costs as lines $5-6$. After getting $R''$, the server recovers the cost accumulation result $\llbracket I \rrbracket$ for each path as line $7$. Moreover, the server takes $-\llbracket I \rrbracket$ and $\textbf{0}$ as the input of $\mathtt{PackObliviousCom}$ to transform path cost-based evaluation into polynomial-based evaluation as line $8$. It is worth noting that the value on the correct path is $0$, and the values of other paths are negative. Therefore, after executing $\mathtt{PackObliviousCom}$, the value on the correct path is $1$, and the values of other paths are $0$. By multiplying the weight vector, only the correctly indexed weight is preserved. The correctness of $\mathtt{PackPathEva}$ is proven in \cref{theoremThree}.

\section{Kangaroo for Large-Scale Evaluation}
\label{epdti}
In this section, we present the inference protocol of Kangaroo under the client-server model for large-scale evaluation, such as random forest. Then, we analyze its computational complexity and compare it with the existing schemes.

\subsection{The Inference Protocol for Random Forests}
When dealing with large-scale models, such as random forests or a deep and large decision tree, the number of decision nodes increases significantly. Kangaroo are capable of handling these models, and we use random forests as an example. {Due to space limitation, the inference protocol for random forests is shown in~\cref{OnlineInferenceCS}.} First, the server performs $\mathtt{FeatureSelPack}$ to select and merge relevant features for multiple tree models. $\mathtt{FeatureSelPack}$ is optimized by $\mathtt{I}$-$\mathtt{PackFeatureSel}$ and is mentioned in~\cref{PackFeatureSel1}. Then, the server and the client jointly execute $\mathtt{PackObliviousCom}$ to complete the comparison. To protect the model structure, the tree model is hiding by introducing dummy nodes and randomly swapping the internal nodes. Therefore, the computations need to be adjusted to eliminate the influence introduced by these modifications, as highlighted in blue. First, compute $R_\gamma * \Upsilon_\gamma^{pack}$ to determine whether the comparison result of each node needs to be flipped. Second, since the true weights are on the left child of the dummy node, the comparison result will be $0$. Third, if the left and right child nodes of the dummy node have been swapped, the comparison result will be $1$. Based on three rules, the comparison results can be recovered. Next, $\mathtt{PackPathEva}$ is executed to obtain the encrypted evaluation results $\{\llbracket T_\gamma \rrbracket\}_{\gamma=1}^\Gamma$. It is worth noting that each ciphertext vector packs the results of $M$ models. Therefore, for each $\mathtt{PackPathEva}$, the client and the server jointly build $M$ tree structures using the structural indices in plaintext, perform path cost-based evaluations over plaintext, and then pack the evaluation results into a single ciphertext vector. Subsequently, $\mathtt{PackObliviousCom}$ is employed to convert all evaluation outcomes into polynomial-based evaluations, and $\llbracket T_\gamma \rrbracket$ is calculated. To respond the result, the server generates a random mask vector $T'$ and sends $\sum_{\gamma=1}^\Gamma \llbracket T_\gamma \rrbracket + \textbf{T}'$, $\sum_{i = 1}^{(\tau^*+1)M} T'[i]$ to the client. The client decrypts and obtains the inference result $\pi = \sum_{i = 1}^{(\tau^*+1)M}T''[i] - \sum_{i = 1}^{(\tau^*+1)M} T'[i] $.


\subsection{Complexity Analysis}
We conduct a detailed analysis and comparison of the online operation costs for two-party inference schemes across two categories of inference: depth-round~\cite{MaT0C21,BaiSCCR22} and constant-round schemes~\cite{Cong00P22,MahdaviNLK23,TaiMZC17,KissNLAS19}. The comparative results are summarized in~\cref{Complexity}. It is evident that depth-round schemes incur a significant increase in communication rounds as the model depth $D$ grows, leading to considerable communication cost in real-world networks. In contrast, constant-round protocols are unaffected by communication latency. However, their performance is heavily influenced by the number of tree nodes $\tau$ and the number of trees $K$, making them unsuitable for evaluating large-scale random forests. Through careful design, our Kangaroo framework significantly reduces the aforementioned computation cost by leveraging efficient amortization. Moreover, Kangaroo can further achieve full amortization through the adaptive encoding adjustment strategy mentioned in~\cref{adaptiveEncoding}.

\section{Enhancement For practical Applications}
To enhance practicality, we introduce several optimization strategies, including same-sharing-for-same-model, latency-aware strategy, and adaptive encoding adjustment.
\begin{figure*} 
    \centering
    \includegraphics[width=0.95\linewidth]{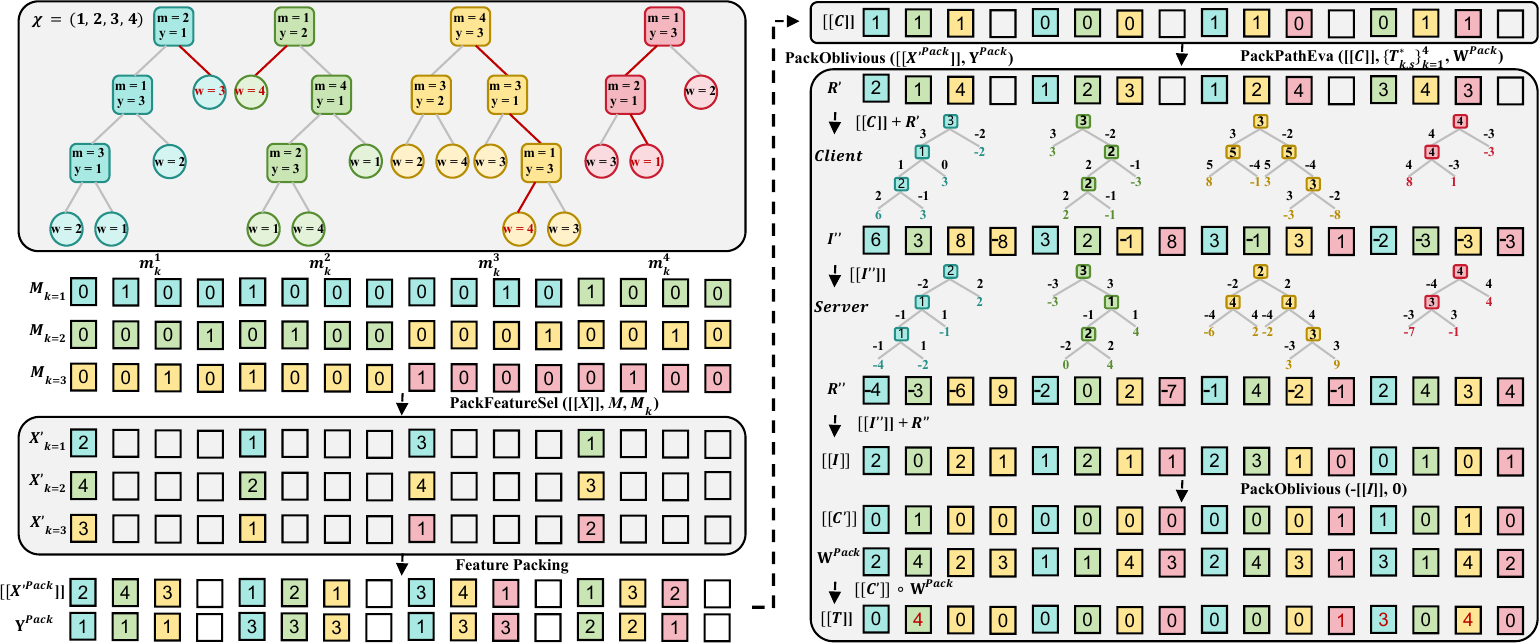}
    \caption{Toy example without model hiding for adaptive encoding adjustment, where $\mathtt{N} = 16$, $M=4$, and $K=4$.}
    \label{ExampleForForest}
\end{figure*}

\subsection{Same-Sharing-for-Same-Model}
\label{samesharing}
In Kangaroo, the server executes the model hiding and extraction to obtain $\{\textbf{M}_k\}_{k=1}^K$, $\{\mathcal{T}^*_{k,s}\}_{k=1}^K$, and $\{\textbf{Y}_{\gamma}^{pack}, \textbf{W}_{\gamma}^{pack}, \Upsilon_{\gamma}^{pack}, \Psi_{\gamma}^{pack}\}_{\gamma=1}^{\Gamma}$ and publishes $\mathcal{T}^*_{k,s}$ to the client. We observe that, due to the obfuscated nature of our $\mathtt{PackPathEva}$, the client cannot distinguish which leaf node is accessed from the paths. Therefore, publishing same model structure indices $\{\mathcal{T}^*_{k,s}\}_{k=1}^K$ can protect the evaluation results while maximizing the entropy of the tree structure~\cite{DuanCLLYL24,09779}, thus preserving the privacy of the original model structures. Additionally, we note that once the packed model is determined, the polynomial length $\mathtt{N}$ and plaintext modulo $q$ are fixed, which also determines the encoder. This implies that the server does not need to perform additional encoding on the packed model when responding to any client request, thereby reducing computation cost. It also ensures that the parameter $\zeta$ is uniquely determined, where we set $\zeta = 2^{\frac{\log q}{2} -1}$ to ensure the correctness. Similarly, same-sharing-for-same-model can be also applicable to outsourced scheme, enhancing its practicality.

\subsection{Latency-Aware Strategy}
In client-server model, the server needs to generate certain parameters to complete the model evaluation. To ensure the protection of sensitive information, these parameters must be randomly generated each time. However, we note that the generation of these parameters, such as $\{\textbf{E},\textbf{A},\textbf{B},\textbf{R},\textbf{R}',\textbf{R}''\}$, is both independent of the model and the client's data, and computationally inexpensive. Therefore, these operations can be amortized and executed during the waiting time for responses from the other party. We refer to this as a latency-aware strategy, which is also applied in other areas. For instance, Kangaroo utilizes encryption operations such as $\llbracket X \rrbracket = \llbracket 0 \rrbracket + \textbf{X}$, where $\llbracket 0 \rrbracket$ can be precomputed during the waiting period. During the process, we also aim to minimize operations on the encoding and ciphertext as much as possible. For example, the server can compute $B\circ R - A \circ R\circ Y$ and encode the result to reduce the number of encodings. When using $(-\llbracket I \rrbracket, \textbf{0})$ as the input of $\mathtt{PackObliviousCom}$, the server can compute $-A \circ R$ and encode it to minimize the negation operations on ciphertext. These optimizations further enhance the practicality. Similarly, the latency-aware strategy can be also applicable to outsourced scheme.

\subsection{Adaptive Encoding Adjustment}
\label{adaptiveEncoding}
In large-scale evaluation, each tree is often pruned to prevent overfitting, resulting in varying structures across different trees~\cite{forestsprune0, forestsprune1, forestsprune2, forestsprune3}. Padding all trees with dummy nodes to ensure that each tree contains exactly $\lceil \frac{\mathtt{N}}{M} \rceil$ nodes would lead to inefficient ciphertext packing and significantly reduce the actual packing utilization of Kangaroo. Fortunately, we can easily adjust the packing size for each tree to solve the problem. This aligns with our design principle of treating each ciphertext slot as a representation of a node. For clarity, we present in~\cref{ExampleForForest} a toy example without applying model hiding. The core idea is to encode features from multiple trees into a single polynomial. During feature selection, the algorithm simultaneously selects node features across multiple trees. The selected features are then further packed and compared. During the path evaluation, it is only necessary to identify which positions correspond to nodes in specific trees. It is evident that the strategy further reduces the number of feature selection and fully exploits every coefficient of the polynomial for inference. Thus, it can achieve the full amortization. The strategy can also be applied to single tree evaluation. It is worth noting that this process does not leak any structural information about the model, as the tree structures are randomly permuted before publishing, thereby meeting our security requirements. Moreover, it is observed that during feature selection, the number of rotations is significantly reduced when $M$ is a power of $2$. Therefore, by adjusting $M$ to $M^*$ and padding with a few dummy features, the inference efficiency can be improved, where $M^*$ is usually a power of $2$. This optimization is also demonstrated in our experiments.

\section{Performance Evaluation}
\label{Performance}
In evaluating Kangaroo within client-server model, we aim to answer the following two main research questions (RQs).

$\bullet$ $\textbf{RQ1}:$ What are the characteristics and advantages of the individual components in Kangaroo compared to recent state-of-the-art (SOTA) methods?

$\bullet$ $\textbf{RQ2}:$ How efficient is Kangaroo in handling large-scale tree models under a WAN setting compared to existing SOTA two-party PDTE schemes?

\subsection{Experimental Setting}
$\bullet$ \textbf{Implementation:}
{The Kangaroo is implemented using Microsoft SEAL version 4.1~\cite{microsoftseal}, which implements the BFV cryptosystem in polynomial and batching technique~\cite{SIMD}.} The data is encrypted with 128-bit security and the default degree of the polynomial $\mathtt{N}$ is set to $8192$. The plaintext modulo $q$ is set to $50$-bit. We provide a clear implementation on GitHub to facilitate verification of the correctness and performance of our schemes. We compare Kangaroo with the SOTA schemes~\cite{MaT0C21,BaiSCCR22, MahdaviNLK23, TaiMZC17, KissNLAS19}. {The comparison is conducted using the following implementations~\cite{secretflow,libOTe,aby,pdte,rasoul}.} {To ensure the reproducibility of our experimental results, all experiments were conducted in a controlled environment with simulated network conditions, unless stated otherwise. We adopt three different bandwidth (bit/second) and round-trip time (RTT) over local-area network (LAN, $1$ Gbps, RTT: $0.1$ ms) metropolitan-area network (MAN, $100$ Mbps, RTT: $6$ ms), and wide-area network (WAN, $40$ Mbps, RTT: $80$ ms) settings to compare their performance.} All experiments are conducted on two ThinkPad-P53 machines with single thread, 23.1 GB RAM, and an Intel Core i5-9400H 2.50GHz processor running on Ubuntu 18.04.6. The experiments show average results, and each is repeated five times independently.

$\bullet$ \textbf{Datasets and Tree Models:}
We utilize datasets from the UCI Machine Learning Repository to verify the correctness of our schemes and evaluate their performance. {Due to space limitation, details of the datasets and tree models are provided in schemes~\cite{KissNLAS19,MahdaviNLK23}. In addition, we also provide detailed model structures and plaintext evaluation interfaces in our GitHub.} In Kangaroo, all tree models are initialized with dummy nodes and randomized. The model is encoded and packed as mentioned in~\cref{workflow,adaptiveEncoding}.

\subsection{Microbenchmarks}
To answer \textbf{RQ1}, we perform a comprehensive set of microbenchmarks to test the efficiency for each of the basic operators in Kangaroo, including secure feature selection, oblivious comparison, and secure path evaluation. 

\begin{figure}
    \centering
    \subfloat[One Feature Selection]{\includegraphics[width=.49\linewidth]{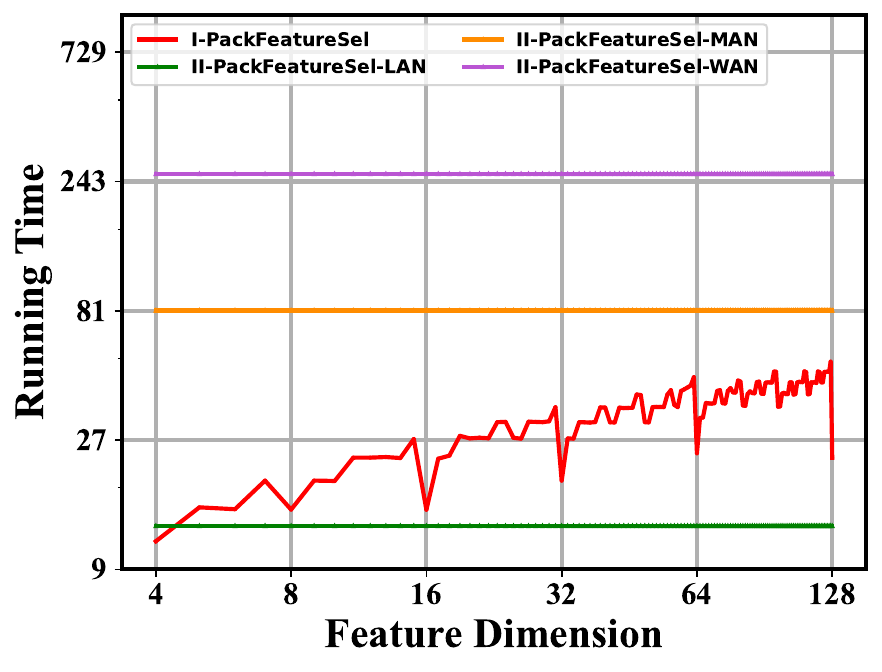}}\hfil
    \subfloat[Fully Feature Selection]{\includegraphics[width=.49\linewidth]{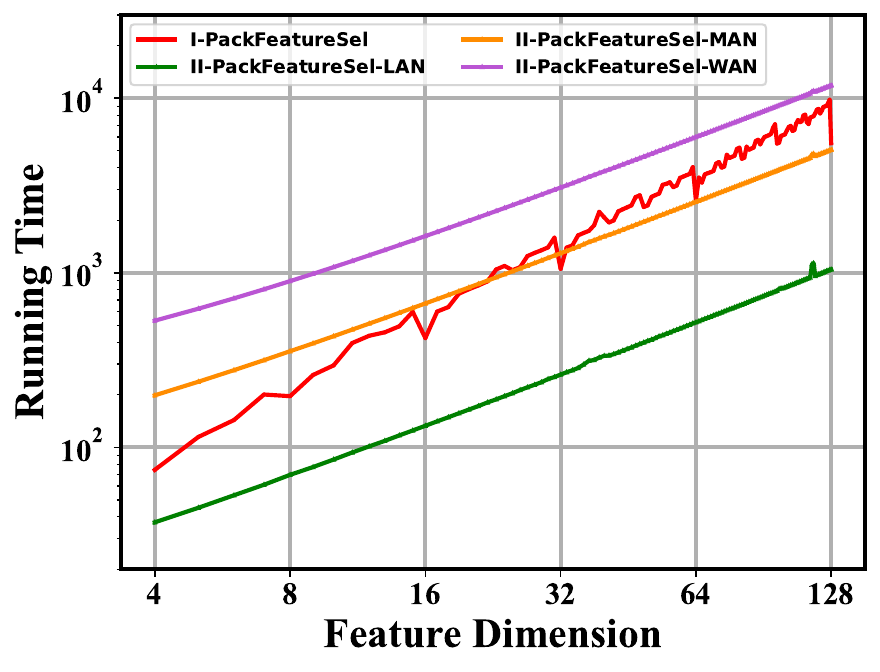}}\hfil
    \caption{The running time (ms) under different network.}
    \label{featureSelectionTime}
\end{figure}

\begin{table}
    \caption{The online running time and communication cost for $1$-out-of-$2$ feature selection.}
    \label{OTcompare}
    \centering
    \begin{tblr}
        {
        colspec         = {*{4}{Q[co=-1,c,m]}Q[co=-1,c,m]},
        columns      = {valign=m},
        cell{1}{1}   = {r=2}{},
        cell{1}{2}   = {c=2}{},
        cell{1}{4}   = {c=2}{},
        rows         = {halign=c},
        row{1}   = {font=\bfseries},
        hline{1-Z}   = {wd=.08em},
        }
                       &  Slient OT~\cite{RaghuramanRT23}  &         & $\mathtt{I}$-$\mathtt{PackFeatureSel}$         &                \\
                       &  $4192$                     &  Amortized    & $4192$       &  Amortized     \\
        Online Time    &  15.00 (ms)                 &  3.66 (us)    &  7.27 (ms)   &  1.77 (us)     \\
        Online Comm.   &  71.54 (KB)                 &  17.89 (B)    &  0           &    0           \\
        \end{tblr}
\end{table}

$\bullet$ $\textbf{Secure Feature Selection:}$ Kangaroo supports two secure feature selection approaches, including $\mathtt{I}$-$\mathtt{PackFeatureSel}$ (\cref{PackFeatureSel}) and $\mathtt{II}$-$\mathtt{PackFeatureSel}$ (\cref{PackFeatureSel2}). The $\mathtt{I}$-$\mathtt{PackFeatureSel}$ enables non-interactive feature selection but incurs a computation cost of $\mathtt{Mul}^* + O(\log M) \mathtt{Rot}$. In contrast, $\mathtt{II}$-$\mathtt{PackFeatureSel}$ scheme requires one interaction with communication cost of $2\cdot\frac{2\mathtt{N} Q}{8192}$ (KB), but only involves $\mathtt{Mul}^* + \mathtt{Dec} + \mathtt{Enc}$ operations. We evaluate the performance of single and fully amortized feature selection, as shown in~\cref{featureSelectionTime}. The fully amortized feature selection is achieved by~\cref{PackFeatureSel1}. From the results, we observe that adjusting $M$ can reduce the number of rotations required. Overall, $\mathtt{I}$-$\mathtt{PackFeatureSel}$ is suitable for scenarios with low feature dimensions and constrained bandwidth, whereas $\mathtt{II}$-$\mathtt{PackFeatureSel}$ performs better in high-bandwidth environments with high feature dimensions. We also compare $\mathtt{I}$-$\mathtt{PackFeatureSel}$ with Slient OT~\cite{RaghuramanRT23} for $1$-out-of-$2$ to finish the feature selection. The result is shown in~\cref{OTcompare}. As shown in~\cref{OTcompare}, $\mathtt{I}$-$\mathtt{PackFeatureSel}$ offers two primary advantages: it is non-interactive and particularly well-suited for small-scale data selection. Moreover, it requires any offline operations, making it lightweight and easy to deploy.

\begin{table}
    \caption{The running time and communication (KB) for oblivious comparison.}
    \label{ComrunningTime}
    \centering
    \begin{tblr}
        {
        colspec         = {*{5}{Q[co=-1,c,m]}Q[co=-1,c,m]},
        columns      = {valign=m},
        rows         = {halign=c},
        row{1}   = {font=\bfseries},
        hline{1-Z}   = {wd=.08em},
        }
                                         &  DGK~\cite{TaiMZC17}    & ABY~\cite{demmler2015aby}  & Cheetah~\cite{LuHZWH23}   & Ours      \\
        Offline Time                     &  0                      & 39.31 (ms)                 & 1363.92 (ms)                  & 0         \\
        Offline Comm.                    &  0                      & 1501.48                    & 1438.32                   & 0         \\
        Amortized Online  Time           &  142.98 \newline (ms)   & 7.66 \newline (us)         & 10.37 \newline (us)        & 0.88 \newline (us) \\
        Amortized Online  Comm.          &  8.25                   & 1.50                       & 0.040                     & 0.10      \\
        Online  Round                    &  1                      & $O(\log t)$                & $O(\log_4 t) $                     & 1      \\
        \end{tblr}
\end{table}

$\bullet$ $\textbf{Oblivious Comparison:}$ As shown in~\cref{ComrunningTime}, Kangaroo achieves the lowest amortized online time for oblivious comparison $\mathtt{PackObliviousCom}$, outperforming DGK~\cite{TaiMZC17}, ABY~\cite{demmler2015aby}, and even the recent Cheetah~\cite{LuHZWH23}. In their scheme, the bit size $t$ is set to $32$-bit. Different from ABY and Cheetah, which require significant offline communication or setup phases, $\mathtt{PackObliviousCom}$ does not rely on offline computation or communication. Furthermore, their communication rounds scale with $O(\log t)$, while $\mathtt{PackObliviousCom}$ requires only a single round of communication. This makes $\mathtt{PackObliviousCom}$ particularly well-suited for real-world, large-scale inference tasks.

\begin{table}
    \caption{The amortized running time and amortized ciphertext size (bit) for different operations.}
    \label{EncryptionrunningTime}
    \centering
    \begin{tblr}
        {
        colspec         = {*{4}{Q[co=-1,c,m]}Q[co=-1,c,m]},
        columns      = {valign=m},
        rows         = {halign=c},
        row{1}   = {font=\bfseries},
        hline{1-Z}   = {wd=.08em},
        }
                                     &  $\mathtt{Enc}$                   & $\mathtt{Add}$                  & $\mathtt{Plain}$-$\mathtt{Mul}$  & $\mathtt{Dec}$      \\
        Paillier ($4096$)            &  5.80 (ms)                        & 3.84 (us)                       &       42.86 (us)                 & 5.80 (ms)              \\
        BFV ($422.18$)               &  0.51 (us) $\textbf{11370}\times$ & 0.08 (us) $\textbf{48}\times$&       0.31 (us) \newline $\textbf{138}\times$              & 0.18 (us)\newline $\textbf{32220}\times$                  \\
        \end{tblr}
\end{table}
$\bullet$ $\textbf{Secure Path Evaluation:}$ During the path evaluation, the path cost-based evaluation is more efficient than polynomial-based evaluation. This is because polynomial-based evaluation requires ciphertexts with a multiplicative depth of $D$, as in the scheme~\cite{ZhaoZXWLL23}, which significantly impacts inference efficiency. In contrast, path cost-based evaluation, such as in~\cite{TaiMZC17}, only involves ciphertext addition and plaintext multiplication. However, as shown in~\cref{Complexity}, this approach still incurs considerable computation cost. While Boolean comparison results can be transformed into arithmetic values to enable path evaluation over non-packed ciphertexts, this method incurs substantial overhead due to the numerous encryption, decryption, addition, and multiplication operations required, all of which grow proportionally with the number of nodes. As illustrated in~\cref{EncryptionrunningTime}, Paillier with $128$-bit security level~\cite{MaHL21} performs poorly in large-scale computations compared to BFV. In Kangaroo, $\mathtt{PackPathEva}$ uses PHE to amortize the costs, resulting in a small number of multiplications, encryptions, and decryptions, as shown in~\cref{Complexity}.

\begin{figure}
    \centering
    \subfloat[Total Communication Cost]{\includegraphics[width=.49\linewidth]{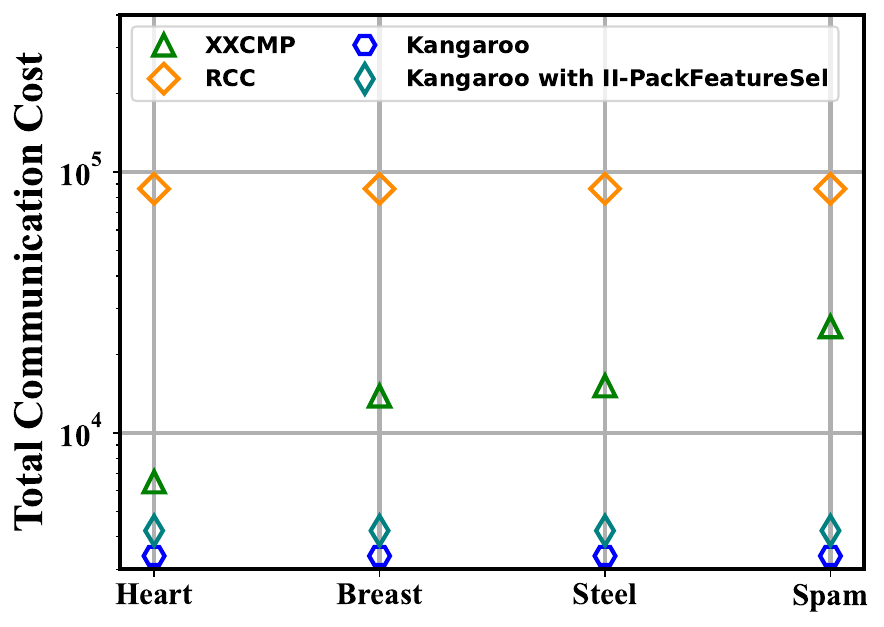}}\hfil
    \subfloat[LAN (1Gbps/0.1ms)]{\includegraphics[width=.49\linewidth]{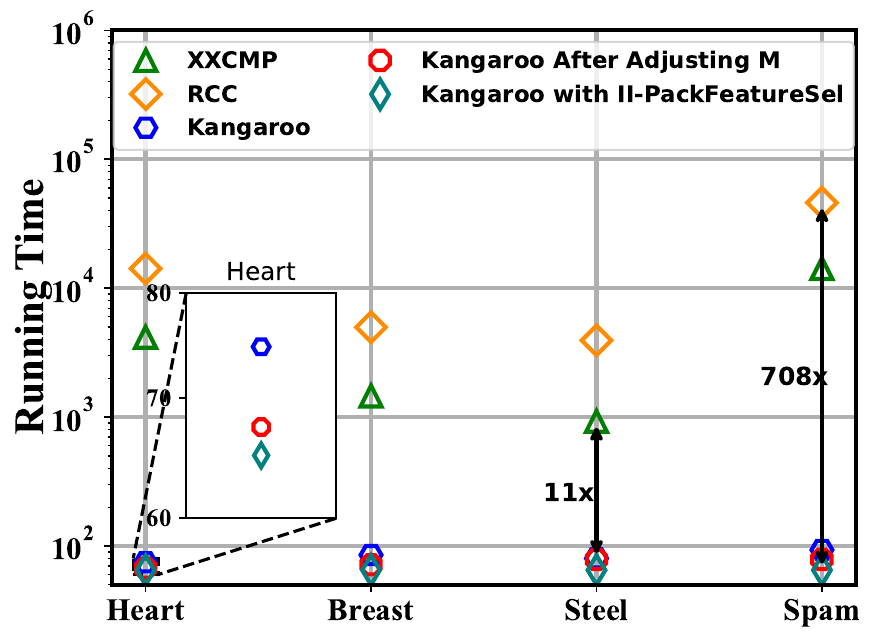}}\hfil
    \subfloat[MAN (100Mbps/6ms)]{\includegraphics[width=.49\linewidth]{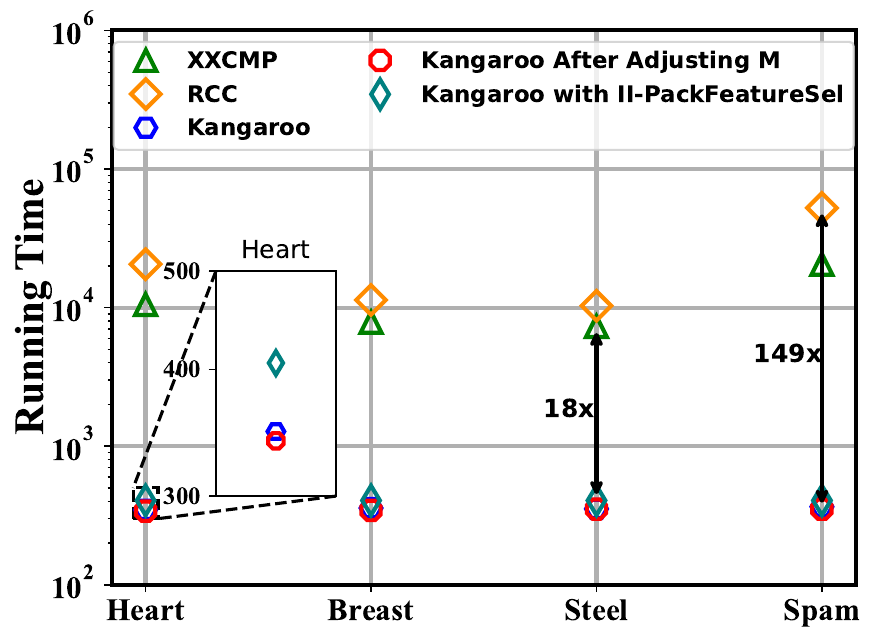}}\hfil
    \subfloat[WAN (40Mbps/80ms)]{\includegraphics[width=.49\linewidth]{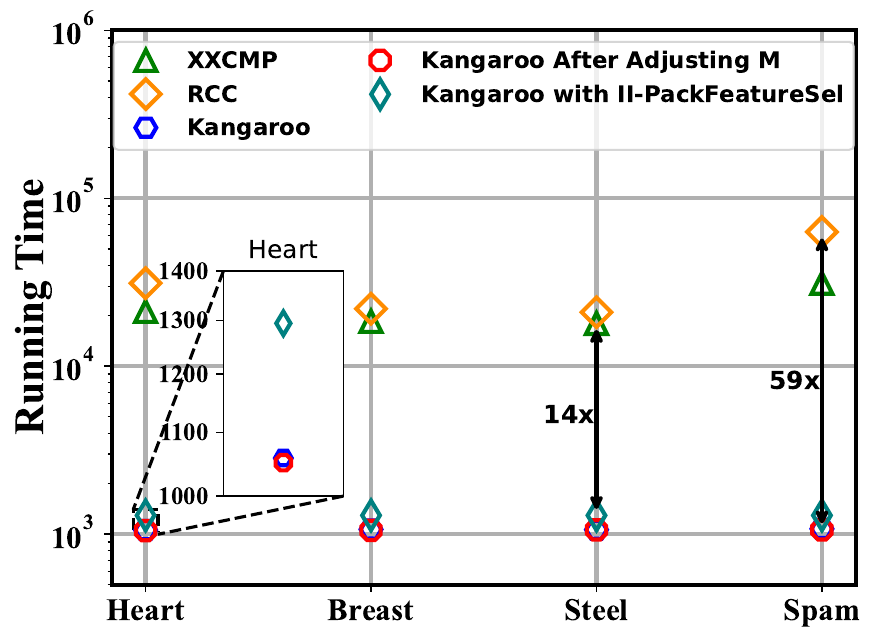}}\hfil
    \caption{{The comparison of total communication cost (KB) and running time (ms) with one round interactive schemes. The $(M, D, T)$ are as follows: Heart $(13, 3, 5)$, Breast $(30, 7, 17)$, Steel $(33, 5, 6)$, and Spam $(57, 16, 58)$.}}
    \label{NonSchemeCompare}
\end{figure}
    
\begin{table*}
    \caption{Total communication cost (KB) and online runtime (ms) for large-scale single tree evaluation in WAN setting. }
    \label{SingleTree}
    \centering
    \small
    \begin{threeparttable}
    \footnotesize
    \begin{tblr}
        {
            width=\textwidth,
            colspec         = {*{12}{Q[co=-1,c,m]}Q[co=-1,c,m]},
            cell{1}{1}   = {r=2}{},
            cell{1}{2}   = {c=2}{},
            cell{1}{4}   = {c=2}{},
            cell{1}{6}   = {c=2}{},
            cell{1}{8}   = {c=2}{},
            cell{1}{10}   = {c=2}{},
            cell{1}{12}   = {c=2}{},
            row{1,2}      = {font=\bfseries},
            hline{3-Y}      = {wd=.08em},
            hline{2}        = {1-Z}{wd=.1em,leftpos=1,rightpos=1,endpos=true},
        }
        \hline                  
        Datasets $(M, D, \tau)$    &  HHH~\cite{TaiMZC17}  &         & HGH~\cite{KissNLAS19}   &         &  GGH~\cite{KissNLAS19} &        &  Sparse~\cite{MaT0C21} &        &  HE-SOS~\cite{BaiSCCR22}  &       & Kangaroo &       \\
                                      &  Comm.      & Time    & Comm.       & Time    &  Comm.      &  Time   &  Comm.      &  Time   &  Comm.      &  Time   & Comm.    &  Time      \\
        \textbf{Digits} $(47, 15, 168)$    &  1163       & 23190   & 1299        & 3578    &  4472       &  3100   &  131        &  3844   &  14079      &  7486   & \textbf{3379}     &  \textbf{1068} ($\textbf{3}\times\sim \textbf{22}\times$)     \\
        \textbf{Diabetes} $(10, 28, 393)$  &  2227       & 43122   & 2973        & 8320    &  9868       &  7525   &  249        &  6614   &  26545      &  13270  & \textbf{3379}     &  \textbf{1050} ($\textbf{6}\times\sim \textbf{41}\times$)      \\
        \textbf{Boston} $(13, 30, 425)$    &  2419       & 46416   & 3157        & 8997    &  10370      &  8138   &  264        &  7058   &  28513      &  14117  & \textbf{3379}     &  \textbf{1050} ($\textbf{7}\times\sim \textbf{44}\times$)      \\
        \hline
    \end{tblr}
    \end{threeparttable}
\end{table*}

Built upon low-latency, efficient, and fully amortized components, Kangaroo is particularly well-suited for large-scale decision tree evaluation in WAN setting. Even when compared with lower-latency schemes~\cite{MahdaviNLK23}, such as one-round communication protocols, RCC and XXCMP, Kangaroo still outperforms them significantly for small-scale decision tree evaluation, achieving up to $14\times$-$59\times$ performance improvements in WAN scenarios. The results are presented in~\cref{NonSchemeCompare}. Clearly, in WAN setting with large-scale decision tree evaluation, Kangaroo will show more superior performance, as its amortization capabilities are fully leveraged under such scenarios. Additionally, we can also observe that adjusting $M$ can improve inference efficiency, such as $M$ from $13$ to $16$ on the Heart dataset in LAN setting.

\subsection{Large-Scale Tree Models Benchmarks} 
To answer the $\textbf{RQ2}$, we evaluate the end-to-end inference latency under WAN setting with large-scale tree models, such as large-scale single tree evaluation and large-scale random forest evaluation, which measures from the time the feature vector is sent until the inference result is received. We also compare Kangaroo with existing two-party inference schemes~\cite{TaiMZC17,KissNLAS19,MaT0C21,BaiSCCR22} to show the efficiency of Kangaroo.

\begin{figure}
    \centering
    \subfloat[$M = 47$]{\includegraphics[width=.49\linewidth]{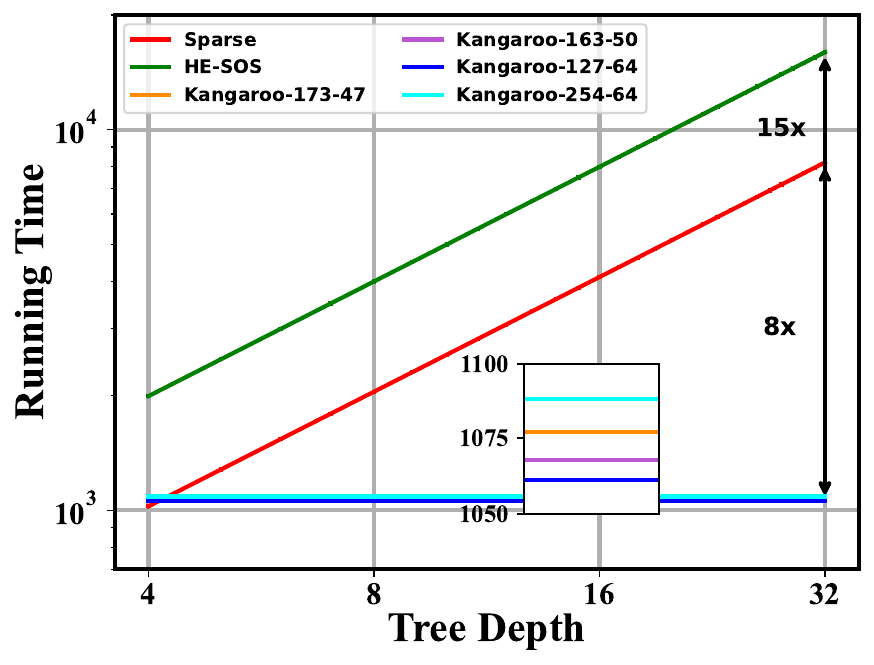}}\hfil
    \subfloat[$M = 13$]{\includegraphics[width=.49\linewidth]{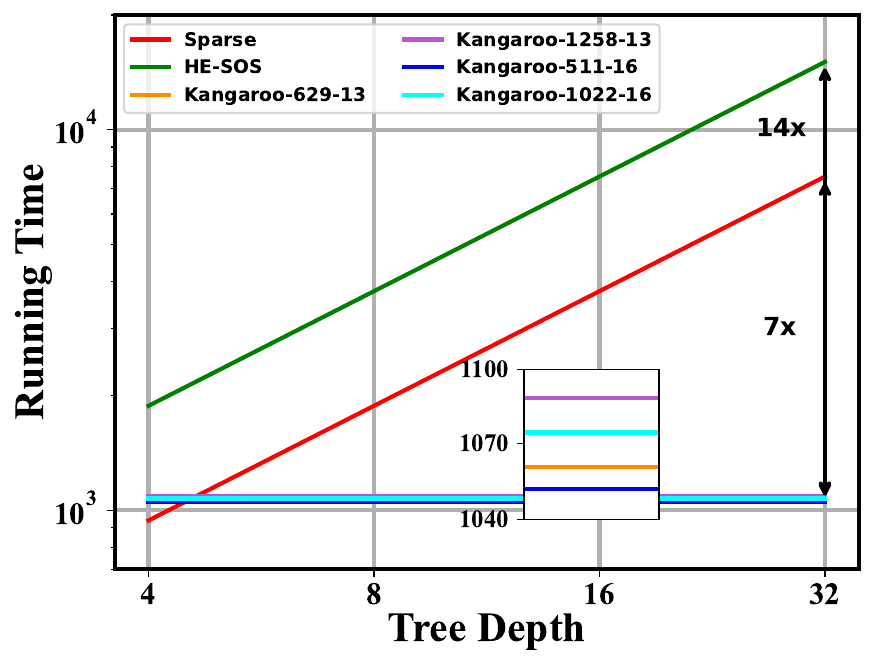}}\hfil
    \caption{The running time (ms) for large-scale single tree evaluation across different tree depths in WAN setting. Kangaroo-$a$-$b$: where $a$ denotes the maximum number of decision nodes supported, and $b$ represents the adjusted feature dimension $M$.}
    \label{RunningTimewithDepth}
\end{figure}

$\bullet$ $\textbf{Large-Scale Single Tree Evaluation:}$  As shown in~\cref{SingleTree}, Kangaroo is highly efficient and well-suited for WAN-based large-scale single tree evaluation. Compared to prior SOTA schemes~\cite{TaiMZC17,KissNLAS19,MaT0C21,BaiSCCR22}, Kangaroo achieves $3\times$-$22\times$ speedup on the Digits dataset, $6\times$-$41\times$ on the Diabetes dataset, and $7\times$-$44\times$ on the Boston dataset, where $M$ is adjusted to $50$, $16$, and $16$. Furthermore, we construct tree models with varying depths and control the number of decision nodes through model pruning. As shown in~\cref{RunningTimewithDepth}, compared with schemes~\cite{MaT0C21, BaiSCCR22}, Kangaroo achieves an improvement of $8\times$-$15\times$ when $M=47$, $D=32$, and $7\times$-$14\times$ when $M=13$, $D=32$. Moreover, Kangaroo can flexibly control the number of decision nodes by adjusting $M$. Furthermore, different from prior schemes~\cite{KissNLAS19, MaT0C21, BaiSCCR22}, Kangaroo operates without any offline computation or communication, making it particularly well-suited for handling large-scale multiple client-side inference requests.

\begin{figure}
    \centering
    \subfloat[Digits]{\includegraphics[width=.49\linewidth]{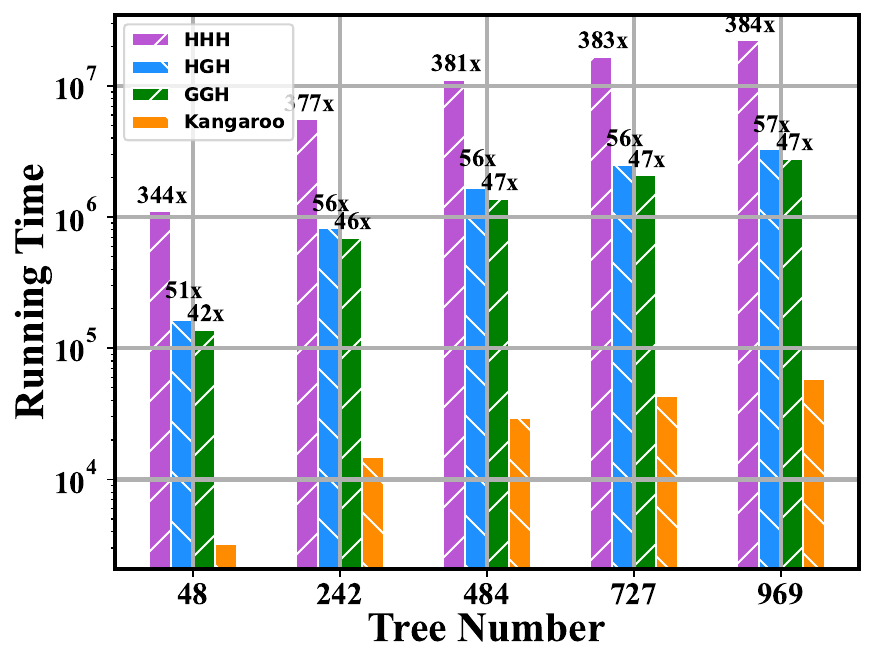}}\hfil
    \subfloat[Boston]{\includegraphics[width=.49\linewidth]{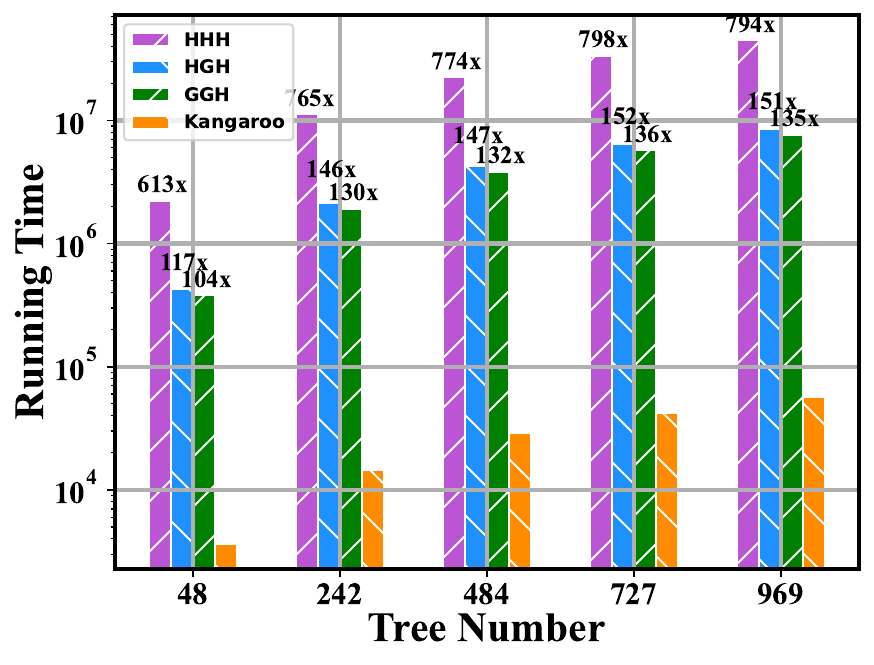}}\hfil
    \caption{The running time (ms) for large-scale random forest evaluation across different tree numbers in WAN setting.}
    \label{RunningTimewithNumber}
\end{figure}

$\bullet$ $\textbf{Large-Scale Random Forest Evaluation:}$ We also construct large-scale random forest by Digits and Diabetes datasets with different tree numbers and compare Kangaroo with the schemes~\cite{TaiMZC17, KissNLAS19}. {The results are shown in~\cref{RunningTimewithNumber}.} When the forest consists of 969 trees, containing 162792 nodes for the Digits dataset and 411825 nodes for the Boston dataset, Kangaroo achieves approximately $47\times$-$384\times$ speedup over existing schemes on Digits, and about $135\times$-$794\times$ improvement on Boston. Moreover, the average inference time per tree is about $60$ ms. All these advantages stem from Kangaroo's full amortization capability, which enables it to consistently achieve the best performance among constant-round schemes.

\begin{table}
    \caption{Total online runtime for privacy-preserving applications under real-world WAN conditions.}
    \label{RealWAN}
    \centering
    \small
    \begin{threeparttable}
    \begin{tblr}
        {
            colspec         = {*{12}{Q[co=-1,c,m]}Q[co=-1,c,m]},
            row{1}      = {font=\bfseries},
            hline{3-Y}      = {wd=.08em},
            hline{2}        = {1-Z}{wd=.1em,leftpos=1,rightpos=1,endpos=true},
        }
        \hline               
        {Applications}   & {JD Cloud (5 Mbps), TP (1 Mbps), RTT (30 ms)}  &  {Ali Cloud (100 Mbps), TP (1 Mbps), RTT (40 ms)} \\
        {Image\\ Recognition}       & 6.96 (s) & 3.18 (s)\\
        {Medical\\ Diagnostics}     & 5.92 (s) & 2.52 (s)\\
        {Financial\\ Forecasting}   & 5.98 (s) & 3.06 (s)\\      
        \hline
    \end{tblr}
    \begin{tablenotes}
        \footnotesize
        \item[1] We leverage the Digits $(47, 15, 168)$, Diabetes $(10, 28, 393)$, and Boston datasets $(13, 30, 425)$ to demonstrate Kangaroo's applicability in image recognition, medical diagnostics, and financial forecasting, respectively. 
    \end{tablenotes}
    \end{threeparttable}
\end{table}

{\subsection{Further Experiments and Discussion}
Kangaroo demonstrates strong performance in enabling large-scale, privacy-preserving decision tree evaluation over WAN environments. It can be seamlessly integrated into a wide range of decision tree-based applications where data privacy is a critical concern. In particular, it is well-suited for scenarios involving sensitive information, such as image recognition, medical diagnostics, and financial forecasting. To demonstrate Kangaroo's capability under real-world WAN conditions, we deploy servers across different cloud platforms. One server is hosted on JD Cloud, equipped with single thread, 3.8 GB RAM, and an Intel(R) Xeon(R) Gold 6148 @ 2.40 GHz processor, running Ubuntu 22.04.5. The other server is deployed on Alibaba Cloud, featuring a single CPU thread, 4 GB RAM, and an Intel(R) Xeon(R) Platinum @ 2.50 GHz processor, also running Ubuntu 22.04.5. The client is still deployed on the ThinkPad-P53 (TP) with single thread. As shown in~\cref{RealWAN}, Kangaroo still demonstrates strong efficiency even under constrained bandwidth conditions. It is worth noting, however, that Kangaroo's performance on small-scale models in LAN settings may be slightly lower than that of schemes specifically optimized for such environments (e.g.,~\cite{MaT0C21}). To improve performance in these scenarios, several optimization strategies, such as reducing the ciphertext modulus and decreasing the polynomial degree, can be employed to further accelerate computation.
}

\section{Conclusion}
\label{conclusions}
In this work, we present Kangaroo, a private and amortized two-party inference framework over WAN for large-scale decision tree evaluation. By utilizing PHE to design a set of new secure feature selection, oblivious comparison, and secure path evaluation protocols, we fully exploit each coefficient in the PHE ciphertexts to efficiently pack decision tree nodes, thereby amortizing both computation and communication costs. Experimental results demonstrate that Kangaroo's core components deliver strong performance, and Kangaroo achieves significant speedups over SOTA two-party PDTE schemes in WAN setting with large-scale models.

{In future work, we plan to build on the design principles of Kangaroo to explore more secure and efficient inference mechanisms, such as those for deep neural networks (DNNs), and develop corresponding optimization techniques to accelerate their performance. We also intend to investigate stronger threat models in outsourced settings, including malicious adversaries, with a focus on ensuring both data privacy and the correctness of inference results.}

\section*{Acknowledgment}
We thank all anonymous reviewers and shepherds for their helpful feedback. This work was supported by National Cryptologic Science Fund of China (2025NCSF02015), National Natural Science Foundation of China (U22B2030, 62572020), Shaanxi Provincial Key Research and Development Program(2024SF2-GJHX-37), Shenzhen Science and Technology Program (JGJZD20240729142310014), Young Elite Scientists Sponsorship Program by CAST (2023QNRC001), and Fundamental Research Funds for the Central Universities (YJSJ25011). Hui Zhu is the corresponding author.

\bibliographystyle{IEEEtran}%
\bibliography{test}



\begin{table}
    \caption{Notations for Kangaroo}
    \label{Notations}
    \centering
    \footnotesize
    \begin{tblr}{
    width        = 1\linewidth,
    colspec      = {lX},
    hline{1,2,Z} = {.08em},
    hline{2}     = {.05em},
    cell{1}{1,2} = {}{cmd=\textbf}
    }
    Notations                                &  Definition                                                                    \\
    $M,D,t$                                  & Feature dimension, tree depth, bit size of feature                             \\
    $K, k$                                   & Total number of tree, $1\leq k\leq K$                                          \\
    $\tau,\tau^*$                            & Total decision node number before and after hiding                             \\
    $\mathcal{T}_{k,s}, \mathcal{T}_{k,s}^*$ & Model structure index vector before and after hiding for $k$-th tree           \\
    $\textbf{m}_k, \textbf{y}_k, \textbf{w}_k,\upsilon_k,\psi_k$ & Feature index, threshold, weight, flip condition, and node status vectors after hiding for $k$-th tree                                                     \\
    $\Gamma, \gamma$                         & Total number after packing, $1 \leq \gamma \leq \Gamma$                        \\
    $\textbf{M}_k$                           & Encoded feature index vector for $k$-th tree                                   \\
    $\textbf{Y}_{\gamma}^{pack}$             & Encoded and packed feature threshold vector                                    \\
    $ \textbf{W}_{\gamma}^{pack}$            & Encoded and packed weight vector                                               \\
    $\Upsilon_{\gamma}^{pack}$               & Encoded and packed flip condition vector                                       \\
    $ \Psi_{\gamma}^{pack}  $                & Encoded and packed node status vector                                          \\
    $\textbf{m}_k^n$                         & $M$-size one-hot vector for $n$-th node of $k$-th tree                  \\ 
    $\mathcal{X}$                            & Client's feature vector $\{x_1,x_2,\cdots, x_M\}$                              \\
    $\mathtt{N},q,Q$                         & Polynomial size, plaintext and ciphertext modulo                               \\
    $\zeta$                                  & Precision parameter                                                            \\
    \end{tblr}
    \end{table}  

\appendices
\label{appendix}

\begin{figure}[!h]
    \begin{tcolorbox}[colback = white, colframe = lightgray]
        \textbf{$\mathtt{PackPathEva}$ Protocol} \\ 
        \textbf{SInput:} The encrypted vector $\llbracket C \rrbracket$, the obfuscated structure index $\mathcal{T}_s^*$, and the encoded weight $\textbf{W}$. \\
        \textbf{CInput:} The obfuscated structure index $\mathcal{T}_s^*$. \\
		\textbf{SOutput:} The encrypted evaluation result vector $\llbracket T \rrbracket$. \\
        $\color{gray}\rhd$ $\color{gray}\textit{The server executes:}$ \\
        {\small1:} $R' \leftarrow\{r'_{1},*,\cdots,*, \cdots,r'_{(\tau^*-1)M+1},*,\cdots,*\}$, $\llbracket I' \rrbracket \leftarrow \llbracket C \rrbracket + \textbf{R}' \Rightarrow$ the client. \\
        $\color{gray}\rhd$ $\color{gray}\textit{The client constructs a tree by}$ $\color{gray}\mathcal{T}_s^*$ $\color{gray}\textit{and executes:}$ \\
        {\small2:} $I' \leftarrow \mathtt{Dec}(\llbracket I' \rrbracket, \mathtt{s}) $, the $I'[(n-1)M+1] \leftarrow n$-th $\mathtt{node}$\texttt{->}$\mathtt{left.cost}$ and the $1- I'[(n-1)M+1] \leftarrow n$-th $\mathtt{node}$\texttt{->}$\mathtt{right.cost}$ for $1\leq n \leq \tau^*$.  \\
        {\small3:} $I'' \leftarrow \{i''_1, 0, \cdots, 0,\cdots, i''_{\tau^*M+1}, 0, \cdots, 0\}$, where $I''[(n-1)M+1]$ is the sum of cost along the $n$-th path and $1\leq n \leq \tau^* +1$. \\
        {\small4:} $\llbracket I''\rrbracket \leftarrow \mathtt{Enc}(I'', \mathtt{pk} ) \Rightarrow$ the server. \\
        $\color{gray}\rhd$ $\color{gray}\textit{The server constructs a tree by}$ $\color{gray}\mathcal{T}_s^*$ $\color{gray}\textit{and executes:}$
        {\small5:} The $-R'[(n-1)M+1] \leftarrow n$-th $\mathtt{node}$\texttt{->}$\mathtt{left.cost}$ and the $R'[(n-1)M+1] \leftarrow n$-th $\mathtt{node}$\texttt{->}$\mathtt{right.cost}$ for $1\leq n \leq \tau^*$. \\
        {\small6:} $R''\leftarrow \{r''_1,0, \cdots, 0,\cdots, r''_{\tau^*M+1},0, \cdots, 0\}$, where $R''[(n-1)M+1]$ is the sum of cost along the $n$-th path and $1\leq n \leq \tau^* +1$. \\
        {\small7:} The server gets the encrypted results of path cost-based evaluation $\llbracket I \rrbracket \leftarrow \llbracket I'' \rrbracket + \textbf{R}''$. \\
        $\color{gray}\rhd$ $\color{gray}\textit{The server and client jointly execute:}$\\
        {\small8:} The server gets $\llbracket C' \rrbracket \leftarrow \mathtt{PackObliviousCom}(-\llbracket I \rrbracket, \textbf{0}) $. 
        \Comment{\\Transform path cost-based into polynomial-based.} \\
        $\color{gray}\rhd$ $\color{gray}\textit{The server executes:}$ \\
        {\small9:} The server gets $\llbracket T \rrbracket \leftarrow \llbracket C' \rrbracket\circ \textbf{W}$. \\
    \end{tcolorbox}
    \caption{Packed path evaluation protocol for single tree.}
    \label{PackPath}
\end{figure}

\begin{figure}[!h]
    \begin{tcolorbox}[colback = white, colframe = lightgray]
        \textbf{The Inference Protocol for Random Forests} \\ 
        \textbf{SInput:} $\{M, \textbf{Y}_{\gamma}^{pack}, \textbf{W}_{\gamma}^{pack}, \Upsilon_{\gamma}^{pack}, \Psi_{\gamma}^{pack}\}$, $\{\textbf{M}_k\}_{k=1}^K$, and $\{\mathcal{T}_{k,s}^*\}_{k=1}^K$.\\
        \textbf{CInput:} $\llbracket X \rrbracket$ and $\{\mathcal{T}_{k,s}^*\}_{k=1}^K$. \\
		\textbf{COutput:} The aggregated evaluation result $\pi$. \\
        $\textbf{1) Secure Feature Selection}$ \\
        $\color{gray}\rhd$ $\color{gray}\textit{The server executes:}$ \\
        {\small1:} $\{\llbracket X_{\gamma}^{'pack} \rrbracket\}_{\gamma}^{pack} \leftarrow \mathtt{FeatureSelPack}(\llbracket X \rrbracket, M, \\\{\textbf{M}_k\}_{k=1}^K)$.\\
        $\textbf{2) Oblivious Comparison}$ \\
        $\color{gray}\rhd$ $\color{gray}\textit{The server and client jointly execute:}$ \\
        {\small2:} $\llbracket V'_\gamma \rrbracket \leftarrow $ the steps $1-5$ of $\mathtt{PackObliviousCom}\\(\llbracket X_{\gamma}^{'pack} \rrbracket, \textbf{Y}_{\gamma}^{pack})$ for $1\leq \gamma\leq \Gamma$. \\
        $\color{gray}\rhd$ $\color{gray}\textit{The server executes:}$ \\
        {\small3:} {\color{blue}$R_\gamma \leftarrow R_\gamma * \Upsilon_\gamma^{pack}$, $R_{\gamma}[i] \leftarrow 0$ if $\Psi_{\gamma}^{pack}[i] = 0$ for $1\leq \gamma\leq \Gamma$ and $1\leq i\leq \tau^*M$.} \\
        {\small4:} In $C'_\gamma$, $C'_\gamma[i] \leftarrow 1$ if $R_\gamma[i] = -1$ or {\color{blue}$(\Upsilon_\gamma^{pack}[i] = - 1$ and $\Psi_{\gamma}^{pack}[i] = 0)$}, $0$ otherwise for $1\leq i\leq \tau^*M$. \\
        {\small5:} The server gets $\llbracket C_\gamma \rrbracket \leftarrow \textbf{C}'_\gamma +  \textbf{R}_\gamma  \circ \llbracket V'_\gamma \rrbracket$. \\
        $\textbf{3) Secure Path Evaluation}$ \\
        $\color{gray}\rhd$ $\color{gray}\textit{The server and client jointly execute:}$ \\
        {\small6:} The server gets $\llbracket T_\gamma \rrbracket \leftarrow \mathtt{PackPathEva}(\llbracket C_\gamma \rrbracket,\\ \{\mathcal{T}_{k,s}^*\}_{k = (\gamma -1)M+1}^{\gamma M}, \textbf{W}_{\gamma}^{pack})$ for $1\leq \gamma\leq \Gamma$. \\
        $\textbf{4) Result Response}$ \\
        $\color{gray}\rhd$ $\color{gray}\textit{The server executes:}$ \\
        {\small7:} $T' \leftarrow \mathcal{Z}_q^{(\tau^*+1)M}$, $\sum_{\gamma=1}^\Gamma \llbracket T_\gamma \rrbracket + \textbf{T}'$, $\sum_{i = 1}^{(\tau^*+1)M} T'[i]\Rightarrow$ the client. \\
        $\color{gray}\rhd$ $\color{gray}\textit{The client executes:}$ \\
        {\small8:} $T'' \leftarrow \mathtt{Dec}(\sum_{\gamma=1}^\Gamma \llbracket T_\gamma \rrbracket + \textbf{T}', \mathtt{s})$.\\
        {\small9:} The client gets $\pi \leftarrow \sum_{i = 1}^{(\tau^*+1)M}( T''[i] -  T'[i]) $.\\
    \end{tcolorbox}
    \caption{The inference protocol for random forests under client-server model.}
    \label{OnlineInferenceCS}
\end{figure}

\section{Kangaroo Extensions}
In this section, we introduce some extensions of Kangaroo, including feature selection and outsourcing setting.

\subsection{Extensions for Packed Feature Selection} 
We propose a feature packing algorithm $\mathtt{FeatureSelPack}$ in~\cref{PackFeatureSel1} to select and merge the feature for multiple trees. Moreover, we propose $\mathtt{II}$-$\mathtt{PackFeatureSel}$ protocol~\cref{PackFeatureSel2} for the LAN setting to perform feature selection, which avoids the rotation overhead introduced in~\cref{PackFeatureSel}. First, the server selects a random vector $E$ and calculates $\llbracket X' \rrbracket = \llbracket X \rrbracket\circ \textbf{M} + \textbf{E}$. Then, the server sends $\llbracket X' \rrbracket$ to the client. After receiving $\llbracket X' \rrbracket$, the client decrypts and then divides it into $\tau^*$ blocks, each containing $M$ values. Next, the client sums the elements within each block to obtain the vector $X''$, encrypts it, and sends it to the server, where each summation result is placed in the corresponding position. Similarly, the server utilizes $E$ to perform summation and then negates each result to obtain the vector $E'$. At last, the server calculates $\llbracket X'' \rrbracket + \textbf{E}'$ to get the selected encrypted vector $\llbracket X' \rrbracket$. The protocol can also be easily extended to support multiple trees to achieve the fully amortized feature selection.

\begin{algorithm}[!h]
	\renewcommand{\algorithmicrequire}{\textbf{SInput:}}
	\renewcommand{\algorithmicensure}{\textbf{SOutput:}}
	\caption{$\mathtt{FeatureSelPack}$}
	\label{PackFeatureSel1}
	\begin{algorithmic}[1]
        \Require $\llbracket X \rrbracket$, $M$, and $\{ \textbf{M}_{k} \}_{k=1}^K$.
		\Ensure $\{\llbracket X_{\gamma}^{'pack}\}_{\gamma =1 }^\Gamma$.
        \LComment{The server executes:}
        \State $\llbracket X'_k \rrbracket \leftarrow \mathtt{I}$-$\mathtt{PackFeatureSel}(\llbracket X \rrbracket, M, \textbf{M}_{k})$ for $1\leq k\leq K$. 
        \State $E \leftarrow \{1, 0, \cdots, 0, \cdots, 1, 0, \cdots, 0\}$. 
        \State The server gets $\llbracket X_{\gamma}^{'pack} \rrbracket \leftarrow \sum_{m=1}^M \mathtt{Rot}(\llbracket X'_{(\gamma-1)M+m} \rrbracket \circ \textbf{E}, m-1)$ for $1\leq \gamma\leq \Gamma$. 
	\end{algorithmic}   
\end{algorithm}

\begin{figure}[t]
    \begin{tcolorbox}[colback = white, colframe = lightgray]
        \textbf{$\mathtt{II}$-$\mathtt{PackFeatureSel}$ Protocol} \\ 
        \textbf{SInput:} $\llbracket X \rrbracket$, $M$, and $\textbf{M}$.    \\
		\textbf{SOutput:} The selected encrypted vector $\llbracket X' \rrbracket$. \\
        $\color{gray}\rhd$ $\color{gray}\textit{The server executes:}$ \\
        {\small1:} $E \leftarrow \mathcal{Z}_q^\mathtt{N}$, $\llbracket X'\rrbracket \leftarrow \llbracket X \rrbracket \circ \textbf{M} +  \textbf{E} \Rightarrow$ the client. \\
        $\color{gray}\rhd$ $\color{gray}\textit{The client executes:}$ \\
        {\small2:} $X' \leftarrow \mathtt{Dec}(\llbracket X'\rrbracket, \mathtt{s})$, $X'' \leftarrow \{\sum_{i=1}^M X'[i], 0, \cdots, 0,\\ \cdots,$ $ \sum_{i=1+M(\tau^*-1)}^{M\tau^*}X'[i],0,\cdots,0\}$. \\
        {\small3:} $\llbracket X''\rrbracket \leftarrow \mathtt{Enc}(X'', \mathtt{pk}) \Rightarrow$ the server. \\
        $\color{gray}\rhd$ $\color{gray}\textit{The server executes:}$ \\
        {\small4:} $E' \leftarrow \{-\sum_{i=1}^M E[i], 0, \cdots, 0, \cdots, -\sum_{i=1+M(\tau^*-1)}^{M\tau^*} \\E[i],$ $0,\cdots,0\}$. \\
        {\small5:} The server gets $\llbracket X' \rrbracket \leftarrow \llbracket X'' \rrbracket + \textbf{E}'$. \\
        \end{tcolorbox}
    \caption{Packed feature selection protocol for single tree.}
    \label{PackFeatureSel2}
\end{figure}


\subsection{Kangaroo for Outsourcing}
\label{outsourceKangaroo}
In this section, we introduce our outsourced extension scheme. It allows us to outsource the heavy computational tasks to a cloud server provider (CSP), while the client and server perform only lightweight operations. In the scheme, the server plays the role of the client and CSP plays the role of the server on the client-server scheme. We assume that CSP will not collude with either the server or the client. First, the server needs to initialize a BFV encryption scheme and generates the public key to finish evaluation services. Then, the server encrypts the $\{\{ \textbf{Y}_{\gamma}^{pack} \}_{\gamma=1}^{\Gamma}, \{ \textbf{W}_{\gamma}^{pack} \}_{\gamma=1}^{\Gamma}, \{ \Psi_{\gamma}^{pack} \}_{\gamma=1}^{\Gamma}, \{ \textbf{M}_{k} \}_{k=1}^K\}$ to get the encrypted models. Next, the server publishes the public keys, $\{\mathcal{T}^*_{k,s}\}_{k=1}^K$, and encrypted models to CSP. The model outsourcing is finished, and the same-sharing-for-same-model strategy is applied to reduce computation cost. When a user requests the service, it encrypts its input $\mathbf{X}$, generates an encoded random vector $\mathbf{T'} \in \mathcal{Z}_q^{(\tau^*+1)M}$, and sends them to the CSP. Subsequently, CSP and the server jointly execute~\cref{OnlineInferenceOut} to obtain the inference result for the user. It is worth noting that, during the comparison phase, to mitigate the impact of node swapping, the server flips the comparison result whenever the children of a node are swapped. To mitigate the impact of dummy nodes, CSP calculates $\textbf{C}'_\gamma \leftarrow \textbf{C}'_\gamma \circ \llbracket \Psi{\gamma}^{pack} \rrbracket$ and $\textbf{R}_\gamma \leftarrow \textbf{R}_\gamma \circ \llbracket \Psi{\gamma}^{pack} \rrbracket$ to ensure that the comparison results for dummy nodes is $0$.


\begin{figure}[!h]
    \begin{tcolorbox}[colback = white, colframe = lightgray]
        \textbf{The Outsourced Inference Protocol} \\ 
        \textbf{CSPInput:} $\{\{\llbracket Y_{\gamma}^{pack} \rrbracket\}_{\gamma=1}^{\Gamma}, \{\llbracket W_{\gamma}^{pack} \rrbracket\}_{\gamma=1}^{\Gamma},\\ \{\llbracket \Psi{\gamma}^{pack} \rrbracket\}_{\gamma=1}^{\Gamma}, \{\llbracket M_{k} \rrbracket\}_{k=1}^K\}$, $M$, and $\{\mathcal{T}_{k,s}^*\}_{k=1}^K$.\\
        \textbf{SInput:} $\{\mathcal{T}_{k,s}^*\}_{k=1}^K$, $\{ \Upsilon_{\gamma}^{pack} \}_{\gamma=1}^{\Gamma}$. \\
        \textbf{CInput:} $\llbracket X \rrbracket$ and $\mathbf{T'}$. \\
		\textbf{COutput:} The aggregated evaluation result $\pi$. \\
        $\color{gray}\rhd$ $\color{gray}\textit{The user executes:}$ \\
        {\small1:} $\llbracket X \rrbracket,\mathbf{T'}$ $\Rightarrow$ CSP. \\
        $\textbf{1) Secure Feature Selection}$ \\
        $\color{gray}\rhd$ $\color{gray}\textit{The CSP executes:}$ \\
        {\small1:} $\{\llbracket X_{\gamma}^{'pack} \rrbracket\}_{\gamma}^{pack} \leftarrow \mathtt{FeatureSelPack}(\llbracket X \rrbracket, M, \\\{\llbracket M_{k} \rrbracket\}_{k=1}^K)$.\\
        $\textbf{2) Oblivious Comparison}$ \\
        $\color{gray}\rhd$ $\color{gray}\textit{The CSP and server jointly execute:}$ \\
        {\small2:} $ V'_\gamma \leftarrow $ the steps $1-4$ of $\mathtt{PackObliviousCom}\\(\llbracket X_{\gamma}^{'pack} \rrbracket, \llbracket  {Y}_{\gamma}^{pack} \rrbracket)$ for $1\leq \gamma\leq \Gamma$. \\
        $\color{gray}\rhd$ $\color{gray}\textit{The server executes:}$ \\
        {\small3:} {\color{blue}$V'_\gamma[i] \leftarrow 1 - V'_\gamma[i]$ if $\Upsilon_{\gamma}^{pack}[i] = -1$}, $\llbracket V'_\gamma \rrbracket \Rightarrow$ CSP for $1 \leq i \leq \tau^*M$ and $1\leq \gamma\leq \Gamma$.\\
        $\color{gray}\rhd$ $\color{gray}\textit{The CSP executes:}$ \\
        {\small4:} In $C'_\gamma$, $C'_\gamma[i] \leftarrow 1$ if $R_\gamma[i] = -1$, $0$ otherwise for $1\leq i\leq \tau^*M$, {\color{blue}$\textbf{C}'_\gamma \leftarrow \textbf{C}'_\gamma \circ \llbracket \Psi{\gamma}^{pack} \rrbracket$ and $\textbf{R}_\gamma \leftarrow \textbf{R}_\gamma \circ \llbracket \Psi{\gamma}^{pack} \rrbracket$}.  \\
        {\small5:} The CSP gets $\llbracket C_\gamma \rrbracket \leftarrow \textbf{C}'_\gamma +  \textbf{R}_\gamma  \circ \llbracket V'_\gamma \rrbracket$. \\
        $\textbf{3) Secure Path Evaluation}$ \\
        $\color{gray}\rhd$ $\color{gray}\textit{The CSP and server jointly execute:}$ \\
        {\small6:} The CSP gets $\llbracket T_\gamma \rrbracket \leftarrow \mathtt{PackPathEva}(\llbracket C_\gamma \rrbracket,\\ \{\mathcal{T}_{k,s}^*\}_{k = (\gamma -1)M+1}^{\gamma M}, \llbracket W_{\gamma}^{pack} \rrbracket)$ for $1\leq \gamma\leq \Gamma$. \\
        $\textbf{4) Result Response}$ \\
        $\color{gray}\rhd$ $\color{gray}\textit{The CSP executes:}$ \\
        {\small7:} $\sum_{\gamma=1}^\Gamma \llbracket T_\gamma \rrbracket + \textbf{T}'\Rightarrow$ the server. \\
        $\color{gray}\rhd$ $\color{gray}\textit{The server executes:}$ \\
        {\small8:} $\sum_{i = 1}^{(\tau^*+1)M}T'' \leftarrow \sum_{i = 1}^{(\tau^*+1)M}\mathtt{Dec}(\sum_{\gamma=1}^\Gamma \llbracket T_\gamma \rrbracket + \textbf{T}', \mathtt{s})\Rightarrow$ the client.\\
        {\small9:} The client gets $\pi \leftarrow \sum_{i = 1}^{(\tau^*+1)M}( T''[i] -  T'[i]) $.\\
    \end{tcolorbox}
    \caption{Outsourced inference protocol for random forests.}
    \label{OnlineInferenceOut}
\end{figure}

\section{Correctness Analysis}
The correctness of the Kangaroo framework depends on three core building blocks. Hence, it suffices to prove the correctness of these components.

\subsection{The Correctness of $\mathtt{PackFeatureSel}$}

\begin{theorem}
    \label{theoremOne}
    After executing $\mathtt{PackFeatureSel}$, the selected encrypted result $\llbracket X' \rrbracket = [[x_{m[1]}],[*],\cdots,[*],\cdots,$ $[x_{m[(\tau^*-1)M+1]}],[*],\cdots,[*]]$ can be obtained.
\end{theorem}

\begin{figure} 
    \centering
    \includegraphics[width=0.8\linewidth]{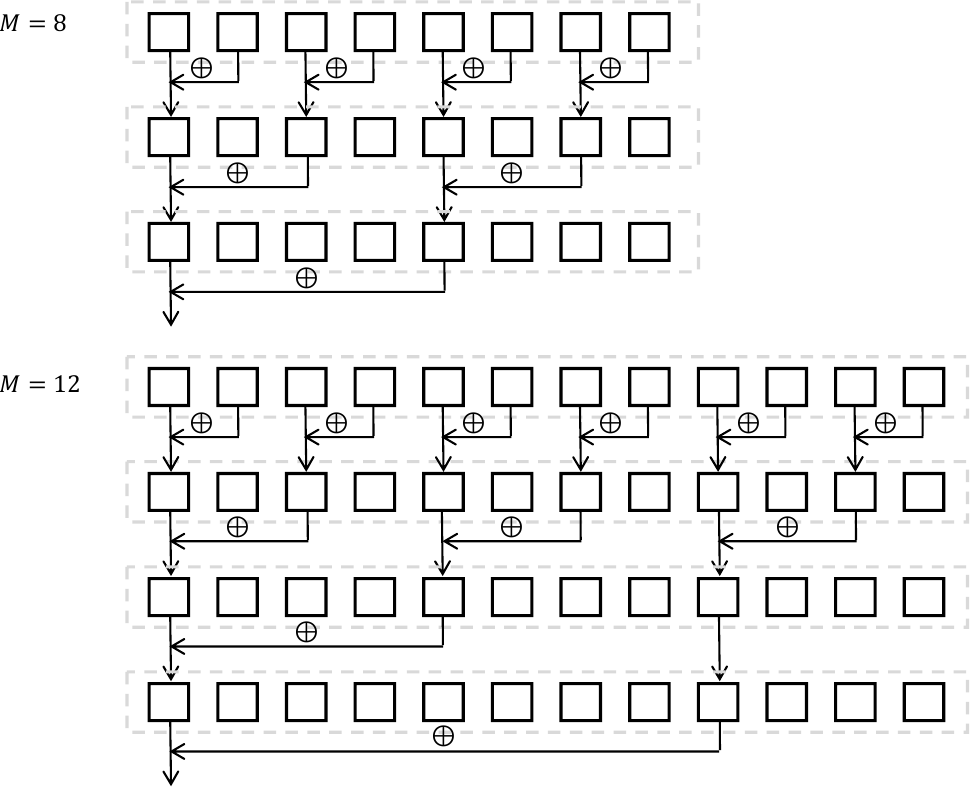}
    \caption{A toy example of $\mathtt{I}$-$\mathtt{PackFeatureSel}$ for $M = 8$ and $M = 12$.}
    \label{featureSelection}
\end{figure}

\textit{Proof.}
To make it easier to understand, we focus only on the first $M$ values in the packed ciphertext. Since the operation is parallel, it is correct for the other positions. First, in $\mathtt{PackFeatureSel}$, $[x_{m[1]}]$ is selected by a one-hot vector $\textbf{m}^1_k$, and other elements are set to $0$. Therefore, the feature selection can be treated as the summation operation. In $\mathtt{I}$-$\mathtt{PackFeatureSel}$, we utilize rotate-and-sum technique to finish the summation of $M$ values. To improve the efficiency, we utilize the division-based approach to finish the summation and the core idea is to sum two adjacent results and store the sum in the previous result. Finally, by iterating, the sum of $M$ values is accumulated into the first position. If a value does not have an adjacent result, it waits to enter the next round. We also show a toy example for $M = 8$ and $M = 12$ in \cref{featureSelection} to better understand our $\mathtt{I}$-$\mathtt{PackFeatureSel}$. In $\mathtt{II}$-$\mathtt{PackFeatureSel}$, the server calculates $-\sum_{m=1}^M(E[m])$, and the client calculates $\sum_{m=1}^M(X'[m] + E[m])$. When the server calculates $\llbracket X'' \rrbracket + \textbf{E}'$, the sum of the $M$ value $\llbracket \sum_{m=1}^M(X'[m]) \rrbracket = \llbracket \sum_{m=1}^M(X'[m] + E[m]) \rrbracket  - \sum_{m=1}^M(\textbf{E}[m])$ is placed in the first position. Thus, \cref{theoremOne} is correct.
$\hfill \square$

\begin{table}
\caption{The Truth Table for Oblivious Comparison.}
\label{trueTable}
\centering
\begin{tblr}{
colspec      = {*{6}{Q[c,m]}},
hline{1,2,Z} = {.1em},
hline{2}     = {.1em},
cell{1}{1,2,3} = {}{cmd=\textbf}
}
\hline
$X'[i]-Y[i]$              &  $R[i]$        &   $V[i]$          &  $V'[i]$ &  $C'[i]$  &  $C[i]$    \\
$   < 0$                &  $-1$         &   $> 0$          &  $1$    &  $1$     &  $0$      \\
$   < 0$                &  $ 1$         &   $< 0$          &  $0$    &  $0$     &  $0$      \\
$\geq 0$                &  $-1$         &   $< 0$          &  $0$    &  $1$     &  $1$      \\
$\geq 0$                &  $1 $         &   $> 0$          &  $1$    &  $0$     &  $1$      \\
\hline
\end{tblr}
\end{table}

\subsection{The Correctness of $\mathtt{PackObliviousCom}$}
\begin{theorem}
    \label{theoremTwo}
    After executing $\mathtt{PackObliviousCom}$, the comparison result $\llbracket C \rrbracket$ of $\llbracket X' \rrbracket$ and $\textbf{Y}$ can be obtained, where $C[i] = 0$ if $X'[i]-Y[i] < 0$, $C[i] = 1$ otherwise.
\end{theorem}

\textit{Proof.}
To make it easier to understand, we focus only on the one value in the packed ciphertext. It is important to note that both $\llbracket X' \rrbracket$ and $\textbf{Y}$ in $\mathtt{PackObliviousCom}$ are quantized, ensuring that $X'[i]$ and $Y[i]$ lie within the range $[0, \zeta]$, where $\zeta = 2^{\frac{\log q}{2} - 1}$. It follows  that $X'[i]-Y[i] \in [-\zeta, \zeta]$. For $A$, $B$, and $R$, we have $\zeta > A[i] > B[i] > 0$ and $R[i]\in\{-1,1\}$. Thus, $A[i]\cdot (X'[i] - Y[i]) +B[i] \in (-\zeta^2 +\zeta,0)\cup (0, \zeta^2+\zeta) = (-2^{\log q -2} + 2^{\frac{\log q}{2} - 1},0)\cup  (0,2^{\log q -2} +2^{\frac{\log q}{2} - 1}) \in (-2^{\log q -2} - 2^{\log q -2},0) \cup (0, 2^{\log q -1} + 2^{\log q -1}) = (-\frac{q}{2},0)\cup (0, \frac{q}{2})$. Clearly, when the sign of $A[i] \cdot (X'[i] - Y[i]) + B[i]$ changes, its value still remains within the range $(-\frac{q}{2}, 0) \cup (0, \frac{q}{2})$. This ensures that the sign of $A[i] \cdot (X'[i] - Y[i]) +B[i] \mod p$ is only affected by $X'[i] - Y[i]$ and $R[i]$. Because $X'[i]-Y[i]$ and $R[i]$ are random, we have them in four cases: 1) $X'[i]-Y[i] < 0$, 2) $X'[i]-Y[i] \geq 0$, 3) $R[i] = -1 $, and 4) $R[i]  = 1$. We show the true table in \cref{trueTable}. It is evident that when $X'[i]-Y[i] < 0$, $C[i] = 0$, and when $X'[i]-Y[i] \geq 0$, $C[i] = 1$. By leveraging SIMD technique, all comparison results satisfy this condition. Therefore, \cref{theoremTwo} is correct.
$\hfill \square$

\subsection{The Correctness of $\mathtt{PackPathEva}$}
\begin{theorem}
    \label{theoremThree}
    After executing $\mathtt{PackPathEva}$, the encrypted evaluation result vector $\llbracket T \rrbracket$, where one element in $T$ is the actual weight value, and the remaining elements are $0$. 
\end{theorem}

\textit{Proof.}
Similarly, we take the case of one node as an example. In the original setup, the left cost of the $i$-th node is set to $C[i]$, while the right cost of $i$-th node is set to $1-C[i]$. Now, on the client side, the left cost of the $i$-th node is set to $C[i] + R'[i]$ and the right cost of $i$-th node is set to $1 - C[i] - R'[i]$. On the server side, the left cost of the $i$-th node is set to $- R'[i]$ and the right cost of $i$-th node is set to $R'[i]$. When the costs corresponding to the client and server are added together, the left and right costs of the $i$-th node are recovered. Therefore, when summing along each path, only one path will yield a sum of $0$, while the sums of the other paths will be greater than $0$. Next, we utilize $\mathtt{PackObliviousCom}$ to convert the path cost-based results into polynomial-based results. It is obvious that taking $(-\llbracket I \rrbracket,0)$ as the input of $\mathtt{PackObliviousCom}$ can finish the operation. This ensures that the result at the position of the real weight is $1$, while the results at all other weight positions are $0$, which means that the encrypted evaluation result vector $\llbracket T \rrbracket$ can be obtained. Therefore, \cref{theoremThree} is correct.
$\hfill \square$

\section{Security Analysis}
{In model inference, it is generally unnecessary to assume an adversary that actively tampers with model correctness. If a client behaves maliciously, any resulting inaccuracies are their own responsibility. Conversely, if a service provider intentionally modifies inference results, it may degrade the user experience, erode user trust, and ultimately damage the provider's reputation and commercial value. Therefore, adopting the semi-honest assumption is reasonable in the scenario. Based on this assumption, we give the security definition and the corresponding security analysis of Kangaroo.}

\subsection{Security Definition}
\label{secureDefinition}
We use the security definition of simulation-based real/ideal worlds model~\cite{Foundations} for two-party computation (2PC). In the real/ideal worlds, there are two probabilistic polynomial time (PPT) adversaries $\mathcal{A}_1$ and $\mathcal{A}_2$ that corrupt party A and party B. The real world is consistent with our system model. All messages that $\mathcal{A}_1$ and $\mathcal{A}_2$ can view are the same as that they can view in our scenario. In the ideal world, there is a simulator $\mathcal{S}_1$ with $\{\mathcal{L}_{1,2}, \mathcal{L}_1\}$ and a simulator $\mathcal{S}_2$ with $\{\mathcal{L}_{1,2}, \mathcal{L}_2\}$, where $\mathcal{L}_{1,2}$ is leakage of our scheme to party A and party B, $\mathcal{L}_1$ leakage of our scheme to party A, and $\mathcal{L}_2$ leakage of our scheme to party B. The simulator $\mathcal{S}_1$ (resp. $\mathcal{S}_2$) will simulate the messages that $\mathcal{A}_1$ (resp. $\mathcal{A}_2$) views in the ideal world. If $\mathcal{A}_1$ and $\mathcal{A}_2$ can only distinguish between the real and ideal worlds with negligible probability, then the private data \ding{172} - \ding{180} remains protected, and our scheme is secure under the leakages ${\mathcal{L}_{1,2}, \mathcal{L}_1, \mathcal{L}_2}$. 

\subsection{The Security of Kangaroo for Client-Server Model}
\label{secure1}
In client-server model, party A is the client and party B is the server. We give the leakages to the client and server.

$\bullet$ Leakages to both client and server. $\mathcal{L}_{1,2}$ includes 1) the public parameters of cryptosystem: $\mathtt{pp}$; 2) the public key: $\{\mathtt{(b,a)}\}$; 3) the feature dimension: $M$; 4) the obfuscated model structure indices: $\{\mathcal{T}^*_{k,s}\}_{k=1}^K$; 5) the precision parameter $\zeta$.

$\bullet$ Leakages to client only. $\mathcal{L}_{1}$ includes the private key $(1,\mathtt{s})$.

$\bullet$ Leakages to server only. $\mathcal{L}_{2}$ includes 1) the packed model $\{\textbf{Y}_{\gamma}^{pack}, \textbf{W}_{\gamma}^{pack}, \Upsilon_{\gamma}^{pack}, \Psi_{\gamma}^{pack}\}$; 2) the feature indices $\{\textbf{M}_k\}_{k=1}^K$.

Based on the above leakages, we construct an ideal world model with two adversaries $\{\mathcal{A}_1, \mathcal{A}_2\}$, a simulator $\mathcal{S}_1$ with $\{\mathcal{L}_{1,2}, \mathcal{L}_1\}$ and a simulator $\mathcal{S}_2$ with $\{\mathcal{L}_{1,2}, \mathcal{L}_2\}$. Because our Kangaroo consists of multiple phases, we simulate them to analyze their security and the security of Kangaroo.

\textbf{1) Simulation of $\mathtt{PackFeatureSel}$.} $\mathtt{PackFeatureSel}$ provides two modes to finish the feature selection. Because $\mathtt{I}$-$\mathtt{PackFeatureSel}$ is non-interactive, we only simulate the process of $\mathtt{II}$-$\mathtt{PackFeatureSel}$ for single tree.

\begin{itemize}
    \item [$\bullet$] $\textbf{Step 1:}$ The server calculates $\llbracket X'\rrbracket =\llbracket X \rrbracket \circ \textbf{M} +  \textbf{E}$ and sends it to the client. As a result, $\mathcal{S}_{1}$ chooses the random vectors ${X'}_{sim}$ by the plaintext space in $\mathtt{pp}$, encrypts them and sends them to $\mathcal{A}_{1}$.

    \item [$\bullet$] $\textbf{Step 2:}$ The client sends an encrypted vector $\llbracket X'' \rrbracket$ to CSP. As a result, $\mathcal{S}_{2}$ encrypts a random vector ${X''}_{sim}$ by $\mathtt{pp}$ and sends it to $\mathcal{A}_{2}$.
\end{itemize}

\textbf{2) Simulation of $\mathtt{PackObliviousCom}$.} During the comparison, there exist one round of communication between client and server. Therefore, the simulations include two steps.

\begin{itemize}
    \item [$\bullet$] $\textbf{Step 1:}$ The server calculates $\llbracket V_\gamma \rrbracket = \textbf{A}_\gamma\circ \textbf{R}_\gamma \circ (\llbracket X'_\gamma \rrbracket - \textbf{Y}_\gamma) + \textbf{B}_\gamma\circ \textbf{R}_\gamma$ and sends it to the client. As a result, $\mathcal{S}_{1}$ chooses a random vector ${V}_{sim}$ by $\zeta$ in $ (-\zeta^2-\zeta,0)\cup (0, \zeta^2 +\zeta)$, encrypts it and sends it to $\mathcal{A}_{1}$.

    \item [$\bullet$] $\textbf{Step 2:}$ The client sends an encrypted vector $\llbracket V'_\gamma \rrbracket$ to the server. As a result, $\mathcal{S}_{2}$ encrypts a random vector ${V'}_{sim}$ by $\mathtt{pp}$ and sends it to $\mathcal{A}_{2}$.
\end{itemize}

\textbf{3) Simulation of $\mathtt{PackPathEva}$.} During the path evaluation, there exist four rounds of communication between client and server as \cref{PackPath} (line $2$, line $6$, and line $12$). Therefore, the simulations include four steps.

\begin{itemize}
    \item [$\bullet$] $\textbf{Step 1:}$ The server calculates $\llbracket I'_\gamma \rrbracket = \llbracket C_\gamma \rrbracket + \textbf{R}'_\gamma$ and sends it to the client. As a result, $\mathcal{S}_{1}$ chooses a random vector ${I'_{sim}}$ by the plaintext space in $\mathtt{pp}$, encrypts it and sends it to $\mathcal{A}_{1}$.
    
    \item [$\bullet$] $\textbf{Step 2:}$ The client sends an encrypted vector $\llbracket I''_\gamma \rrbracket$ to the server. As a result, $\mathcal{S}_{2}$ encrypts a random vector ${I''}_{sim}$ by $\mathtt{pp}$ and sends it to $\mathcal{A}_{2}$.
    
    \item [$\bullet$] $\textbf{Step 3:}$ It is same as $\textbf{Step 1}$ of the simulation of $\mathtt{PackObliviousCom}$.
    
    \item [$\bullet$] $\textbf{Step 4:}$ It is same as $\textbf{Step 2}$ of the simulation of $\mathtt{PackObliviousCom}$.
\end{itemize}

\textbf{4) Simulation of Result Response.} During the result response, there exist one round of communication between client and server. Therefore, the simulations include one step.

\begin{itemize}
    \item [$\bullet$] $\textbf{Step 1:}$ The server sends $\sum_{\gamma=1}^\Gamma \llbracket T_\gamma \rrbracket + \textbf{T}'$ and $\sum_{i=1}^{\mathtt{N}}T'[i]$ to the client. As a result, $\mathcal{S}_{1}$ chooses a random vector ${T''_{sim}}$ and a random value $T'_{sim}$, encrypts ${T''_{sim}}$ and sends $\llbracket T''_{sim} \rrbracket, T'_{sim}$ to $\mathcal{A}_{1}$.
\end{itemize}

Based on the real and ideal worlds, we formally define the security of Kangaroo.

\begin{tcolorbox}[colback = white, colframe = lightgray]
\begin{Define}
    [Security of Kangaroo]
    \label{define1}
    The Kangaroo is selectively secure iff for any two PPT adversaries $\{\mathcal{A}_1,\mathcal{A}_2\}$, there exist two simulators $\{\mathcal{S}_1, \mathcal{S}_2\}$ with $\{\mathcal{L}_{1,2}, \mathcal{L}_1, \mathcal{L}_2\}$ to simulate an ideal world such that $\{\mathcal{A}_1,\mathcal{A}_2\}$ distinguish the views from the real world or the ideal world with negligible probability.
\end{Define}
\end{tcolorbox}

\begin{theorem}
    Kangaroo is selectively secure with the $\{\mathcal{L}_{1,2},\mathcal{L}_{1},\mathcal{L}_{2}\}$.
\end{theorem}

\textit{Proof.} Based on the simulation-based real/ideal worlds model, we analyze the security of each phase. During the $\textbf{Step 2}$ in the simulation of $\mathtt{PackFeatureSel}$, the client sends the encrypted vector $\llbracket X'' \rrbracket$ to the server. Since the ciphertext is encrypted using BFV, which is proven to be IND-CPA secure~\cite{BFV}, the server remains unaware of the encrypted content. When $\mathcal{S}_{2}$ sends an encrypted random vector $\llbracket {X''}_{sim} \rrbracket$ to $\mathcal{A}_{2}$, $\mathcal{A}_{2}$ cannot decrypt the ciphertext and it is also indistinguishable. Therefore, $\mathcal{A}_2$ cannot distinguish the data from the ideal world or the real world. It means that the $\textbf{Step 2}$ in the simulation of $\mathtt{PackFeatureSel}$ is secure and the client's data sent to the server is protected. Similarly, $\textbf{Step 2}$ in the simulation of $\mathtt{PackObliviousCom}$ and $\textbf{Steps 2 and 4}$ in the simulation of $\mathtt{PackPathEva}$ can be proven in the same manner. \ding{176}, \ding{177}, \ding{178}, \ding{179} and \ding{180} are protected from the server.

During the $\textbf{Step 1}$ in the simulation of $\mathtt{PackFeatureSel}$, the server sends blinded encrypted vectors as $\llbracket X' \rrbracket = \llbracket X \rrbracket\circ  \textbf{M}  + \textbf{E}$ to the client. Therefore, $\llbracket X' \rrbracket$ are random. When $\mathcal{S}_1$ sends an encrypted random vector $\llbracket {X'}_{sim} \rrbracket$ to $\mathcal{A}_{1}$, $\mathcal{A}_1$ decrypts it, and it is also random. Therefore, $\mathcal{A}_1$ cannot distinguish the data from the ideal world or the real world. Therefore, $\mathtt{PackFeatureSel}$ is secure, and \ding{172} is protected from the client.

During the $\textbf{Step 1}$ in the simulation of $\mathtt{PackObliviousCom}$, the server sends the encrypted vector $\llbracket V \rrbracket$ to the client. It is noting that $\llbracket V_\gamma \rrbracket = \textbf{A}_\gamma\circ \textbf{R}_\gamma \circ (\llbracket X'_\gamma \rrbracket - \textbf{Y}_\gamma) + \textbf{B}_\gamma\circ \textbf{R}_\gamma$. The client decrypts and gets $V_\gamma = A_\gamma \circ R_\gamma (X'_\gamma - Y_\gamma) + B_\gamma \circ R_\gamma$, where the values of vector $X'_\gamma  - Y_\gamma$ are random by two encoded random vector $A_\gamma$ and $B_\gamma$ and the signs of vector $ X'_\gamma  - Y_\gamma$ are also random by the encoded random vector $R_\gamma$. It means that $V_\gamma$ is random in $(-\zeta^2 - \zeta,0) \cup (0,\zeta^2 + \zeta)$, and the signs are oblivious by observing~ \cref{trueTable}. Note that when $R_\gamma[i] = 1$, the value $V[i]$ lies within the range $(-\zeta^2 + \zeta, 0) \cup (0, \zeta^2 + \zeta)$; whereas when $R_\gamma[i] = -1$, the range becomes $(-\zeta^2 - \zeta, 0) \cup (0, \zeta^2 - \zeta)$. This implies that the sign might be leaked with a probability of only $\frac{2 \zeta +2\zeta}{2(\zeta^2+\zeta)} = \frac{2}{\zeta+1}$, where $\zeta =2^{\frac{\log q}{2} - 1}$. The probability is negligible due to the sufficiently large setting of the modulus $q$. When $\mathcal{S}_{1}$ sends an encrypted random vector $\llbracket {V}_{sim} \rrbracket$ to $\mathcal{A}_{1}$, $\mathcal{A}_1$ decrypts it. The result is also random and its signs are oblivious. Therefore, $\mathcal{A}_1$ can distinguish the data from the ideal world or the real world with negligible probability. It means that the difference between $X'_\gamma$ and $Y_\gamma$, as well as its sign, are protected. Therefore, $\mathtt{PackObliviousCom}$ is oblivious, and \ding{173}, \ding{179}, and \ding{180} are protected from the client. We also give an enhanced packed oblivious comparison protocol as~\cref{PackObliviousCom1}. This requires modifying the inputs to $\llbracket 2 \circ \textbf{X}' \rrbracket$ and $\textbf{2} \circ \textbf{Y} - \textbf{1}$, and constraining them within the range $[0, \zeta]$. It is not difficult to observe that the value of $V$ is bounded within the range $(-\zeta^2 - \zeta, 0) \cup (0, \zeta^2 + \zeta) \in (-\frac{q}{2},0)\cup (0, \frac{q}{2})$. It ensures correctness and randomness while also eliminates the impact of potential that the sign is leaked with negligible probability. Moreover, the incurred overhead is negligible due to our latency-aware strategy.

During the $\textbf{Step 1}$ in the simulation of $\mathtt{PackPathEva}$, the server sends $\llbracket I' \rrbracket$ to the client. It is noting that $\llbracket I'_\gamma \rrbracket = \llbracket C_\gamma \rrbracket + \textbf{R}'_\gamma$, where the vector $C_\gamma$ is randomized by $\textbf{R}'_\gamma$. When $\mathcal{S}_{1}$ sends an encrypted random vector $\llbracket {I'}_{sim} \rrbracket$ to $\mathcal{A}_{1}$, $\mathcal{A}_1$ decrypts it and it is also random. Therefore, $\mathcal{A}_1$ cannot distinguish the data from the ideal world or the real world. It means that the comparison result vector $\llbracket C_\gamma \rrbracket$ is protected. When the client executes the path cost-based evaluations over plaintext, the results are randomized, and the decision path is protected. Throughout the process, the model structure remains randomized, thereby preserving its confidentiality. Additionally, by converting the path cost-based evaluations into polynomial-based evaluations, all non-essential weight information is hidden, and only the weights associated with the true inference path are revealed. Thus, the weight vector is protected. \ding{174}, \ding{175}, \ding{178}, and \ding{179} are protected from the client. 

During the result response, the server sends $\sum_{\gamma=1}^\Gamma \llbracket T_\gamma \rrbracket + \textbf{T}'$ and $\sum_{i=1}^{\mathtt{N}}T'[i]$ to the client. It is noting that $\textbf{T}'$ is random. It means that $\sum_{\gamma=1}^\Gamma \llbracket T_\gamma \rrbracket + \textbf{T}'$ and $\sum_{i=1}^{\mathtt{N}}T'[i]$ are random. When $\mathcal{S}_{1}$ sends $\llbracket {T''_{sim}} \rrbracket$ and $T'_{sim}$ to $\mathcal{A}_{1}$, they are also random. Therefore, $\mathcal{A}_1$ cannot distinguish the data from the ideal world or the real world. \ding{174} and \ding{178} are protected from client. 

In all phases of Kangaroo, both $\mathcal{A}_1$ and $\mathcal{A}_2$ can only distinguish the real and ideal world with negligible probability, and the private data \ding{172} - \ding{180} are protected. Therefore, our Kangaroo is selectively secure with $\{\mathcal{L}_{1,2},\mathcal{L}_{1},\mathcal{L}_{2}\}$ and provides the same security as these schemes~\cite{BostPTG15,TaiMZC17,LuZS18,KissNLAS19,TuenoKK19,Cong00P22,MahdaviNLK23}. $\hfill \square$

\begin{figure}[t]
    \begin{tcolorbox}[colback = white, colframe = lightgray]
        \textbf{$\mathtt{PackObliviousCom}^+$ Protocol} \\ 
        \textbf{SInput:} The encrypted and encoded vectors $\llbracket 2\circ X' \rrbracket$, $\textbf{2}\circ \textbf{Y} -\textbf{1}$, where $\llbracket 2\circ X' \rrbracket$ and $\textbf{2}\circ \textbf{Y} -\textbf{1} \in [0, \zeta]$. \\
		\textbf{SOutput:} The encrypted comparison result $\llbracket C \rrbracket$. \\
        $\color{gray}\rhd$ $\color{gray}\textit{The server executes:}$ \\
        {\small1:} $A \leftarrow \{a_{1},\cdots,a_{\mathtt{N}}\}$, $B\leftarrow \{b_{1},\cdots,b_{\mathtt{N}}\}$, where $\zeta > A[{n}] > B[{n}] > 0$ and $1\leq n\leq \mathtt{N}$. \\
        {\small2:} $R\leftarrow \{r_{1},\cdots,*, r_{\mathtt{N}}\},R'\leftarrow \{r'_{1},\cdots,*, r'_{\mathtt{N}}\}$, where $R[n],R'[n] \leftarrow \{1,-1\}$ by flipping a coin.  \\
        {\small3:} $\llbracket V \rrbracket \leftarrow  \textbf{A}\circ \textbf{R} \circ\textbf{R}' \circ(\textbf{2}(\llbracket X' \rrbracket - \textbf{Y})+\textbf{1}) + \textbf{R} \circ\textbf{B} \Rightarrow$ the client. \\
        $\color{gray}\rhd$ $\color{gray}\textit{The client executes:}$ \\
        {\small4:} $V \leftarrow \mathtt{Dec}(\llbracket V \rrbracket, \mathtt{s}) $, $V' \leftarrow \{v'_{1},\cdots, v'_{\mathtt{N}}\}$, where $V'[n] = 0$ if $V[n] < 0$, $1$ otherwise. \\
        {\small5:} $\llbracket V'\rrbracket \leftarrow \mathtt{Enc}(V', \mathtt{pk} ) \Rightarrow$ the server. \\
        $\color{gray}\rhd$ $\color{gray}\textit{The server executes:}$ \\
        {\small6:} $C' \leftarrow \{c'_{1},\cdots,c'{\mathtt{N}}\}$, where $C'[n] = 1$ if $R[n] = -1$, $0$ otherwise. \\
        {\small7:} The server gets $\llbracket C \rrbracket \leftarrow \textbf{C}' +  \textbf{R}  \circ \llbracket V' \rrbracket$.
    \end{tcolorbox}
    \caption{Enhanced packed oblivious comparison protocol.}
    \label{PackObliviousCom1}
\end{figure}

{$\textit{Remark 2.}$ Based on the security proofs, we conclude that the intermediate results observed by the client are randomized and oblivious. Therefore, the client cannot derive any meaningful information from these intermediate outputs~\cite{araki2016high,WaghGC19}. In addition, to mitigate risks posed by arbitrary or abusive queries, we recommend integrating query-based payment and rate-limiting mechanisms into Kangaroo. These enhancements further strengthen the practicality of Kangaroo and improve its resilience against denial-of-service (DoS) attacks.
	}

\subsection{The Security of Kangaroo for Outsourcing}
\label{secure2}
In outsourcing scenario, party A is the server and party B is CSP. We give the leakages to the server and CSP.

$\bullet$ Leakages to both server and CSP. $\mathcal{L}_{1,2}$ includes 1) the public parameters: $\mathtt{pp}$; 2) the public key: $\{\mathtt{(b,a)}\}$; 3) the feature dimension: $M$; 4) the encrypted model parameters $\{\{\llbracket Y_{\gamma}^{pack} \rrbracket\}_{\gamma=1}^{\Gamma}, \{\llbracket W_{\gamma}^{pack} \rrbracket\}_{\gamma=1}^{\Gamma}, \{\llbracket \Psi{\gamma}^{pack} \rrbracket\}_{\gamma=1}^{\Gamma}, \{\llbracket M_{k} \rrbracket\}_{k=1}^K\}$; 5) the obfuscated model structure indices: $\{\mathcal{T}_{k,s}^*\}_{k=1}^K$. 6) the precision parameter $\zeta$.

$\bullet$ Leakages to server only. $\mathcal{L}_{1}$ includes 1) the private key $(1,\mathtt{s})$, 2) the packed model parameters.

$\bullet$ Leakages to CSP only. $\mathcal{L}_{2}$ includes 1) the encrypted feature vector $\llbracket X \rrbracket$, 2) the encoded random vector $\textbf{T}'$.

In the outsourcing setting, the server takes the role of the client, while the CSP takes on the role of the server as in client-server model. Consequently, the analysis of outsourcing setting follows similarly to that of client-server model, providing the same level of security~\cite{ZhengDWWN22,ZhengWWDN23}.



\end{document}